\documentclass[twoside,twocolumn,9pt]{article}
\usepackage{extsizes}
\usepackage[super,sort&compress,comma]{natbib} 
\usepackage[version=3]{mhchem}
\usepackage[left=1.5cm, right=1.5cm, top=1.785cm, bottom=2.0cm]{geometry}
\usepackage{balance}
\usepackage{mathptmx}
\usepackage{sectsty}
\usepackage{graphicx} 
\usepackage{lastpage}
\usepackage[format=plain,justification=justified,singlelinecheck=false,font={stretch=1.125,small,sf},labelfont=bf,labelsep=space]{caption}
\usepackage{float}
\usepackage{fancyhdr}
\usepackage{fnpos}
\usepackage{amsmath,amssymb}
\usepackage[english]{babel}
\addto{\captionsenglish}{%
  
}
\usepackage{array}
\usepackage{droidsans}
\usepackage{charter}
\usepackage[T1]{fontenc}
\usepackage[usenames,dvipsnames]{xcolor}
\usepackage{setspace}
\usepackage[compact]{titlesec}
\usepackage{hyperref}
\usepackage{gensymb}
\usepackage{subcaption}
\usepackage{upgreek}
\usepackage{placeins}
\usepackage{braket}
\usepackage{bm}
\newcommand{\proton}{$^{1}$H}
\newcommand{\phos}{$^{31}$P}
\newcommand{\deltad}{$\Delta_{d}$}
\newcommand{\probe}{\textit{\textbf{P}}}

\newcommand{\fmoc}{\textit{\textbf{FBTrp}}}
\newcommand{\fmocS}{\textit{\textbf{FBTrp-{S}}}}
\newcommand{\fmocR}{\textit{\textbf{FBTrp-{R}}}}

\newcommand\nnfootnote[1]{%
  \begin{NoHyper}
  \renewcommand\thefootnote{}\footnote{#1}%
  \addtocounter{footnote}{-1}%
  \end{NoHyper}
}

\usepackage{xcolor}

\usepackage{epstopdf}

\definecolor{cream}{RGB}{222,217,201}

\begin{document}

\pagestyle{fancy}
\thispagestyle{plain}
\fancypagestyle{plain}{
\renewcommand{\headrulewidth}{0pt}
}

\makeFNbottom
\makeatletter
\renewcommand\LARGE{\@setfontsize\LARGE{15pt}{17}}
\renewcommand\Large{\@setfontsize\Large{12pt}{14}}
\renewcommand\large{\@setfontsize\large{10pt}{12}}
\renewcommand\footnotesize{\@setfontsize\footnotesize{7pt}{10}}
\makeatother

\renewcommand{\thefootnote}{\fnsymbol{footnote}}
\renewcommand\footnoterule{\vspace*{1pt}%
\color{cream}\hrule width 3.5in height 0.4pt \color{black}\vspace*{5pt}} 
\setcounter{secnumdepth}{5}

\makeatletter 
\renewcommand\@biblabel[1]{#1}            
\renewcommand\@makefntext[1]%
{\noindent\makebox[0pt][r]{\@thefnmark\,}#1}
\makeatother 
\renewcommand{\figurename}{\small{Fig.}~}
\sectionfont{\sffamily\Large}
\subsectionfont{\normalsize}
\subsubsectionfont{\bf}
\setstretch{1.125} 
\setlength{\skip\footins}{0.8cm}
\setlength{\footnotesep}{0.25cm}
\setlength{\jot}{10pt}
\titlespacing*{\section}{0pt}{4pt}{4pt}
\titlespacing*{\subsection}{0pt}{15pt}{1pt}

\fancyhead{}
\renewcommand{\headrulewidth}{0pt} 
\renewcommand{\footrulewidth}{0pt}
\setlength{\arrayrulewidth}{1pt}
\setlength{\columnsep}{6.5mm}
\setlength\bibsep{1pt}

\makeatletter 
\newlength{\figrulesep} 
\setlength{\figrulesep}{0.5\textfloatsep} 

\newcommand{\topfigrule}{\vspace*{-1pt}%
\noindent{\color{cream}\rule[-\figrulesep]{\columnwidth}{1.5pt}} }

\newcommand{\botfigrule}{\vspace*{-2pt}%
\noindent{\color{cream}\rule[\figrulesep]{\columnwidth}{1.5pt}} }

\newcommand{\dblfigrule}{\vspace*{-1pt}%
\noindent{\color{cream}\rule[-\figrulesep]{\textwidth}{1.5pt}} }

\makeatother

\twocolumn[
  \begin{@twocolumnfalse}
\vspace{1em}
\sffamily

\noindent\LARGE
{\textbf{Towards detection of molecular parity violation via chiral co-sensing: the $^1$H/$^{31}$P model system}
}

\vspace{0.3cm} 

\noindent\large{Erik Van Dyke$^{\ast}$\textit{$^{a,b}$}, James Eills\textit{$^{c}$}, Kirill Sheberstov\textit{$^e$}, John Blanchard\textit{$^e$}, Manfred Wagner\textit{$^{f}$}, Robert Graf\textit{$^{f}$}, Andr{\'e}s Emilio Wedenig\textit{$^{g}$}, Konstantin Gaul\textit{$^{b,g}$}, Robert Berger\textit{$^{g}$}, Rudolf Pietschnig\textit{$^{h}$},  Denis Kargin\textit{$^{h}$},
 Danila A. Barskiy$^{\ast}$\textit{$^{a,b}$}, and 
 Dmitry Budker\textit{$^{a,b,i}$}} \\

\noindent\normalsize{

Fundamental weak interactions have been shown to violate parity in both nuclear and atomic systems. However, observation of parity violation in a molecular system has proven an elusive target. Nuclear spin dependent contributions of the weak interaction are expected to result in energetic differences between enantiomers manifesting in nuclear magnetic resonance (NMR) spectra as chemical shift differences on the order of $\upmu$Hz to mHz for high-$Z$ nuclei. By employing simultaneous measurements of the diastereomeric splittings for a light and a heavy nucleus in solution-state NMR, residual chemical shift differences persisting in non-chiral environment between enantiomers of chiral compounds smaller than the typical linewidth of high-field NMR may be resolved. Sources of error must be identified and minimized to verify that the observed effect is, in fact, due to parity violation and not systematic effects. This paper presents a detailed analysis of a system incorporating \textsuperscript{31}P and \textsuperscript{1}H NMR to elucidate the systematic effects and to guide experiments with higher-$Z$ nuclei where molecular parity violation may be resolved.  

}

\end{@twocolumnfalse} \vspace{0.6cm}
]


\renewcommand*\rmdefault{bch}\normalfont\upshape
\rmfamily
\section*{}
\vspace{-1cm}


\nnfootnote{\textit{$^{a}$~Institute for Physics, Johannes Gutenberg University Mainz, 55128 Mainz, Germany.}}
\nnfootnote{\textit{$^{b}$~Helmholtz Institute Mainz, 55128 Mainz, Germany; GSI Helmholtz Center for Heavy Ion Research, Darmstadt, Germany.}}
\nnfootnote{\textit{$^{c}$~Institute of Biological Information Processing (IBI-7), Forschungszentrum Jülich, Jülich 52425, Germany}}
\nnfootnote{\textit{$^{d}$~Quantum Technology Center, University of Maryland, College Park, Maryland, MD, 20742 USA}}
\nnfootnote{\textit{$^{e}$~Laboratoire des biomolécules, LBM, Département de chimie, École normale supérieure, PSL University, Sorbonne Université, CNRS, 75005 Paris, France.}}
\nnfootnote{\textit{$^{f}$~Max Planck Institute for Polymer Research, 55128 Mainz, Germany.}}
\nnfootnote{\textit{$^{g}$~Institute for Chemistry, Philipps University Marburg, 35032 Marburg, Germany.}}
\nnfootnote{\textit{$^{h}$~Institute for Chemistry, University of Kassel, 34132 Kassel, Germany.}}
\nnfootnote{\textit{$^{i}$~Department of Physics, University of California at Berkeley, Berkeley CA 94720, USA.}}

\nnfootnote{\dag~Electronic Supplementary Information (ESI) available: [details of any supplementary information available should be included here]. See DOI: 10.1039/cXCP00000x/}

\nnfootnote{* Corresponding author.}



\section{Introduction}
\subsection{Parity violation in atoms and molecules}
Parity violation (PV) in nuclear weak interactions was first suggested by Lee and Yang \cite{Lee1956} and soon confirmed by Wu and colleagues in beta decay of spin polarized \textsuperscript{60}Co nuclei \cite{Wu1957}. A possibility of parity violation in electron-nucleus interactions in atoms was suggested by Zel'dovich \cite{Zeldovich1959} but was estimated to be too small in simple atoms like hydrogen. The discovery of weak neutral currents in neutrino scattering \cite{HASERT1973121, HASERT1973138, Cline1997} rejuvenated the interest in the detecting PV in atoms and it was at that time that Bouchiat and Bouchiat \cite{Bouchiat1974} realized that PV effects are strongly enhanced in heavy atoms. Observations of atomic PV were subsequently reported by Barkov and Zolotorev \cite{Barkov1978}, Conti et al. \cite{Conti1979} and other groups. Atomic PV experiments have contributed to establishing what is now known as the standard model of particles and interactions and since then has become a field of precision measurement, see, for example, the review \cite{Safronova2018}. 

From the early days of atomic PV, it has been recognized that parity violation should also manifest in molecules \cite{Yamagata1966, Mason1984}; in particular, while PV does not produce first-order energy shifts in nondegenerate states \cite{Khriplovich1991}, there are, in fact, first-order energy shifts in chiral molecules since a state with a fixed chirality is a coherent superposition of opposite-parity states. Somewhat surprisingly to the atomic, molecular, and optical physics community, detecting molecular parity violation remains as a yet unmet challenge, both for chiral and non-chiral systems \cite{berger:2019}.

Among various other manifestations of parity violation in chiral molecules (see Refs.~\cite{quack:1989,quack:2008,schwerdtfeger:2010,berger:2019} for reviews) is the appearance of differences in chemical shift between enantiomers undergoing nuclear magnetic resonance (NMR) \cite{barra:1986,Barra1987,barra:1988a}. Here the magnitude of the effect could reach into the millihertz range for favorable cases \cite{Gorshkov1982,Barra1996,robert:2001,laubender:2003,soncini:2003,weijo:2005,laubender:2006,bast:2006,weijo:2007,weijo:2009,nahrwold:09,nahrwold:2014}. While measurable in principle, the effect is hard to detect in practice. Indeed, performing experiments with separated enantiomers would require reliable reproducibility of experimental parameters such as the magnetic field at parts per trillion level, which is beyond current technology. On the other hand, performing measurements in a mixture of enantiomers would appear impossible because the separation of the spectral lines due to the PV effect would be deep within the NMR linewidth.

\subsection{Detection of PV with diastereomerism}

\begin{figure}
\centering
  \includegraphics[width=3.3in]{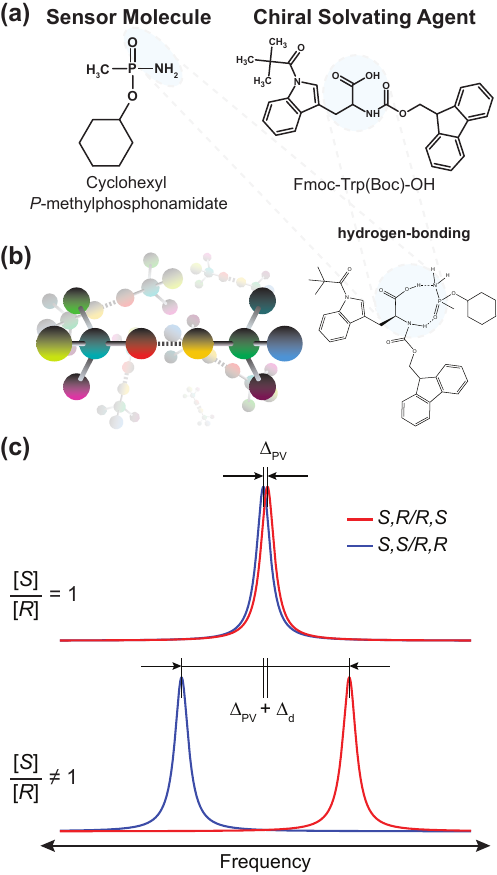}
  \caption{\textbf{(a)} Chemical structures of the chiral sensor molecule and chiral solvating agent used in this work. Hydrogen-bonding interactions generate transient diastereomeric complexes between a racemic mixture of the sensor and a controlled \textit{S}:\textit{R} ratio of the solvating agent. The sensor molecule contains two groups of spin-1/2 nuclei which display variable diastereomeric splitting with respect to enantiomeric ratio of the chiral solvating agent (CSA), and combined measurements of the two function as a co-sensor. \textbf{(b)} The binding interaction between \probe\ and \fmoc\ is expected to occur as hydrogen-bonding at the amino and phosphoryl groups of \probe\ and the amino and carboxylic acid groups of \fmoc. \textbf{(c)} Depiction of NMR spectra of diastereomeric complexes at the racemic point (top) and where an excess of one CSA enantiomer is present (bottom). Diastereomeric splitting (\deltad) appears in addition to splitting caused by parity violation ($\Delta_{\mathrm{PV}})$, which is small relative to the typical NMR linewidth.}
  \label{fgr:figure1}
\end{figure}

A possible solution may be offered by the use of the diastereomerism effect---the splitting of the NMR lines of an enantiomeric mixture of chiral molecules in the presence of a chiral, nonracemic environment \cite{Dale1968, Porto2014}. If one could trace the splitting from the case where the lines are well split in a solvent with one chirality, through the racemic (effectively achiral) solvent, to the opposite-chirality solvent, the PV effect would manifest as nonzero intercept of the splitting. 

The challenge of this approach is finding the exact racemic point of the solvent. In principle, this can be done using precision optical polarimetry techniques, however, it is difficult to do this with sufficient sensitivity and with the necessary control over systematic effects. 

A solution was proposed in Ref.\,\cite{Eills2017}. The idea is that PV effects scale with the atomic number of the nucleus ($Z^a$ with $2>a>5$, see e.g. Ref.~\cite{Gorshkov1982}), so measuring diastereomeric splittings for a heavy and a light nucleus on the same sensor molecule allows using the former as the probe of PV, while the latter as an independent probe of the solvent chirality. This approach is reminiscent of intramolecular comagnetometry used in molecular experiments searching for parity- and time-reversal violating  permanent electric dipole moments \cite{Andreev2018,Roussy2023}. 

In Ref.\,\cite{Eills2017} a proof-of-principle experiment was carried out, where $^{13}$C was used as stand-in for the heavy nucleus, while protons were used as the light nuclei. While the actual PV effect in this system was too small to be detected, that work showed that it was possible to achieve the required sensitivity to energy shifts in the millihertz range, on the order of the size of the effect that could be expected for the heaviest NMR nuclei in chiral molecules with favorable properties.

\subsection{Tunable chiral environment}

It is important to understand that our experiments are possible because we are operating in the regime of rapid chemical exchange. A probe molecule in a solution containing a mixture of the chiral solvating agent of opposite chiralities, in the regime of rapid chemical exchange, samples the opposite chiral environment many times with respect to the $T_2^*$ time scale that determines the effective measurement time width of the spectral lines. This is a regime of strong motional narrowing, in which the diastereomeric complex formed by interactions between the probe molecule and chiral solvating agent is well characterized by its enantiomeric ratio. In particular, the enantiomeric ratio for a racemic mixture of chiral solvating agents (CSA) is 1:1 and no line splitting occurs. Note that, in the opposite limit of slow chemical exchange, even for a racemic solvent, one would observe diastereomeric line splitting: in the absence of parity violation, the spectral lines for \textit{R,R} and \textit{S,S} probe-solvent combinations would overlap; but split from the overlapping lines for \textit{R,S} and \textit{S,R} pairs.

\subsection{Goals of the present work}

In the context of the search for molecular parity violation, we investigate a chiral system containing a relatively heavy atom, \phos. This work builds upon previous work which demonstrated that using a chiral co-sensing system of diastereomeric complexes allows mitigation of systematic errors \cite{Eills2017}, and provides a more in-depth look at the sources of error in such a system. 

Here, a chiral solvating agent (CSA) is used to generate diastereomeric complexes at a lower concentration than previously explored allowing the probing of a heavier spin-1/2 nucleus, \phos, to serve as an intermediate step or alternative pathway towards measuring complexes with high-$Z$ nuclei where PV may be on the order of mHz. Using a CSA as apposed to a chiral solvent has the advantages of allowing a wider range of solvents to be used as well as improved control over concentrations and subsequent binding and dissociation dynamics which are directly related to concentrations of diastereomer forming monomers. 

While \phos\ is still not heavy enough to observe the PV effect, using this heavier nucleus offers an opportunity to explore systematic effects that will be all-important for the choice/synthesis of heavier molecules where detection of PV would be, finally, expected.

\section{Experimental}

\subsection{Choice of the system}
Chiral solvating agents (CSAs) are commonly used in solution-state NMR to resolve mixtures of enantiomers. This is done by dissolving a CSA, typically with one stereogenic center ($S$ or $R$), in a solution containing a target chiral molecule which in turn generates diastereomeric complexes through transient bonding interactions between the CSA and the target molecule \cite{Cefali2024}. In ideal cases, the result of this interaction is one or more nuclei of a target chiral molecule displaying two distinct chemical shifts in NMR spectra corresponding to complexes containing matching (\textit{S,S/R,R}) or opposing (\textit{S,R/R,S}) stereogenic centers. A paper by Li and Raushel \cite{Li2007} detailed such a system where diastereomeric splitting was observed in \proton\ and \phos\ NMR spectra of a chiral phosphorus compound. Following this work we decided to use a chiral phosphonamidate, [amino(methyl)phosphoryl]oxycyclohexane (\probe), and substituted amino acid, Fmoc-(\textit{S})-Trp(Boc)-OH or Fmoc-(\textit{R})-Trp(Boc)-OH (\fmocS and \fmocR), that when combined displayed diastereomeric splitting in both \phos\ and \proton\ spectra of \probe, a key point that makes the sensor molecule \probe\ suitable as a co-magnetometer. We choose the $S$ and $R$ naming convention to assign the absolute configuration of chiral centers as opposed to the L and D convention which has limitations when multiple stereogenic centers are present in a molecule. 

\subsection{Synthesis and NMR characterization of the phosphorus probe molecule}

Cyclohexyl \textit{P}-methylphosphonamidate was prepared according to an adapted literature procedure.\cite{Li2007} 1.270\,g (12.7\,mmol) of dry cyclohexanol were dissolved in 20\,mL of diethyl ether and cooled in an ethanol bath to below –80\,\textdegree C. Dropwise addition of 5.1\,ml (12.7\,mmol, 2.5M in hexanes) \textit{n}-butyllithium yielded a suspension, which was stirred for 5 minutes. Thereafter a solution of 1.688\,g (12.7\,mmol) methylphosphonic dichloride in 40\,mL of diethyl ether was added dropwise at -80\,\textdegree C over the course of 30\,minutes, the mixture was stirred for 30\,minutes at -70\,\textdegree C and for further 30 minutes at room temperature. The solid precipitate was filtered off and at 0\,\textdegree C gaseous ammonia was passed through the clear solution for 3 minutes leading to the formation of a suspension. Volatile compounds were removed under reduced pressure (400\,mbar), the residue was resuspended with 50\,mL of diethyl ether and filtered. Removing of volatile compounds at 400\,mbar followed by recrystallization from diethyl ether yielded the product as a colorless solid in 20\% yield (450\,mg).

$^{1}$H NMR (400\,MHz, CDCl$_{3}$) $\delta$ [ppm]$\colon$4.41 (m, 1H), 2.74 (m, 2H, br), 2.05--1.10 (m, 13H); $^{13}$C NMR (101\,MHz, CDCl$_3$) $\delta$ [ppm]$\colon$73.83 (d, $^2$\textit{J}$_{PC}$ = 6.4 Hz, \textit{C}H), 34.24 (d, $^3$\textit{J}$_{PC}$ = 1.9 Hz), 34.20 (d, $^3$\textit{J}$_{PC}$ = 1.5 Hz), 25.3 (\textit{C}H$_2$), 24.0 (\textit{C}H$_2$), 23.9 (\textit{C}H$_2$),  15.9 (d, $^1$\textit{J}$_{PC}$ = 132.2 Hz, \textit{C}H$_3$); $^{31}$P NMR (202\,MHz, CDCl$_{3}$) $\delta$ [ppm]$\colon$31.7. MS (ESI-HR) $m/z$: 200.0813 ([M+Na]$^{+}$, 27\%) calculated$\colon$200.0816; 377.1714 ([2M+Na]$^{+}$, 100\%), calculated$\colon$377.1735.

Cyclohexyl \textit{N},\textit{N}-diethyl-\textit{P}-methylphosphonamidate was prepared in a similar manner. 1.030\,g (10.3\,mmol) dry cyclohexanol were dissolved in 20 mL of diethyl ether and cooled in an ethanol bath to –80\,\textdegree C. Dropwise addition of 4.1\,mL (10.3\,mmol, 2.5M in hexanes) \textit{n}-butyllithium yielded a suspension, which was stirred for 5 minutes. Thereafter a solution of 1.360\,g (10.2\,mmol) methylphosphonic dichloride in 40\,mL of diethyl ether was added dropwise at –80\,\textdegree C, the mixture stirred for 20 minutes at –80\,\textdegree C and then for another hour at room temperature. Addition of 2.1\,mL (20.6\,mmol, 2 eq) of diethylamine furnished a suspension. The reaction mixture was filtered, the solid washed with 20 mL of diethyl ether and the crude oily product condensed to dryness. Water work-up and extraction with diethyl ether (30\,mL H$_{2}$O and 2x 30\,mL Et$_{2}$O) followed by two-fold vacuum distillation (Bp=80--82\,\textdegree C at 4.5 $\cdot$ 10$^{-2}$\,mbar) yielded an oily product (430\,mg, 17\%, 1.75\,mmol, >95\% purity).
$^{1}$H NMR (500\,MHz, C$_{6}$D$_{6}$) $\delta$ [ppm]$\colon$4.41--4.25 (m, 1H), 2.92 (dq, $^3$\textit{J}$_\mathrm{PH}$=10.7\,Hz, $^3$\textit{J}$_\mathrm{HH}$=7.1\,Hz, 4H), 2.12--1.98 (m, 1H), 1.82--1.69 (m, 1H), 1.62-1.51 (m, 3H), 1.46--1.35 (m, 1H), 1.31--0.99, (m, 8H), 0.89 (t, $^3$\textit{J}$_\mathrm{HH}$= 7.1\,Hz);
$^{13}$C NMR (101\,MHz, C$_{6}$D$_{6}$) $\delta$ [ppm]$\colon$72.3 (d, $^2$\textit{J}$_\mathrm{PC}$=6.5\,Hz), 38.7 (d, $^2$\textit{J}$_\mathrm{PC}$=4.6\,Hz, N\textit{C}H$_2$), 34.5 (d, $^3$\textit{J}$_\mathrm{PC}$=3.0\,Hz, \textit{C}H$_2$), 34.1 (d, $^3$\textit{J}$_\mathrm{PC}$=5.0\,Hz, \textit{C}H$_2$), 25.7 (\textit{C}H$_2$), 24.1 (\textit{C}H$_2$), 23.9 (\textit{C}H$_2$), 14.4 (d, $^3$\textit{J}$_{PC}$=2.0\,Hz, \textit{C}H$_3$), 13.7 (d, $^1$\textit{J}$_\mathrm{PC}$=133.0\,Hz, \textit{C}H$_3$);
$^{31}$P NMR (202\,MHz, C$_{6}$D$_{6}$) $\delta$ [ppm]$\colon$30.2.
MS (APCI-HR) $m/z$: 234.1624 ([M+H]$^{+}$, 10\%) calculated$\colon$234.1623; 152.0849 ([M-Cy+2H]$^{+}$, 100\%), calculated$\colon$ 152.0841.

\subsection{Sample preparation}

All stock solutions were prepared in a glove-box under an atmosphere 
of nitrogen (>99\%). For the first enantiodiscriminatory titration, seperate solutions  of \probe\ and \fmocS\ and \fmocR\ (Merck, Novabiochem) were prepared by dissolving solid analyte in chloroform-d (Eurisotop),  which was used without further purification. Individual samples were prepared by pipetting solutions of either \fmocS\ or \fmocR\ at a concentration of 20\,mM into a  5\,mm NMR tube containing 250\,$\upmu$L of 20\,mM \probe . Equal volumes of \fmoc\ and \probe\ solutions were used in order to reach a final concentration of 10\,mM for both \fmoc\ and \probe\ and the desired ratio of \fmocS:\fmocR. This was repeated to generate solutions of \fmoc\ chirality ranging from 100\% \fmocS\ to 100\% \fmocR\ in steps of 6.25\%. Samples near the racemic point (56.25\% through 43.75\% \fmocS) were omitted due to difficulty distinguishing peaks of diastereomers at these concentrations resulting from substantial spectral overlap.

Samples used in the second enantiodiscriminatory titration (see section \ref{birdSubsection}) were prepared using a slightly modified procedure. Under an atmosphere of nitrogen, 18.7\,mg of \probe\ were dissolved in 10.5\,mL of chloroform-d to generate a 20 mM solution. Using a syringe (Hamilton, 5 mL), this solution was then split between two vials: one containing 20 mM \fmocS\ and the other containing 20 mM \fmocR\ to produce two solutions with a concentration of 10\,mM \fmoc\ (S or R) and 10\,mM \probe. These solutions were then combined directly in 5 mm NMR tubes using a syringe (Hamilton, 1 mL) to generate 0.5\,mL samples with the desired enantiomeric ratio of S:R \fmoc.

\subsection{NMR spectrometer and pulse sequences}
All \proton\ and \phos\ NMR spectra were obtained using an 850 MHz (20.0\,T, Bruker) spectrometer with a QXI 850\,MHz S6 5\,mm multi-nuclei probe with Z gradient at a temperature of 298\,K. \proton\ spectra for the first titration were obtained using a 90\textdegree\ RF pulse with a pre-polarization time of 10 s (methyl \proton\ $T_1$ = 2.4\,s, measured by inversion recovery). \phos\ spectra were acquired using a 90\textdegree\ pulse and inverse gated proton decoupling. Each \phos\ spectrum is the average of 64 transients with 32k points and 0.95\,s of acquisition time and a pre-polarization time of 3\,s (\phos\ $T_1$ = 2.2\,s measured by inversion recovery).

For the second enantiodiscriminatory titration, \proton\ spectra were acquired using a BIRD (bilinear  rotational decoupling) \cite{Garbow1982} pulse sequence to suppress broad resonances belonging to the cyclohexane moiety of \probe\ at frequencies overlapping the methyl-proton peaks from which diastereomeric splitting values were extracted. Parameters for this pulse sequence included a pre-polarization time of 5\,s, an inter pulse delay time of 30.3 ms and a final relaxation delay ($\tau_\mathrm{null}$) of 100 ms before application of a 90\textdegree acquisition pulse (see SI). Both inter pulse delay and $\tau_\mathrm{null}$ were determined empirically using the popt experiment in TopSpin. \phos\ spectra were acquired using a simple 90\textdegree\ pulse scheme without \proton\ decoupling, using a pre-polarization time of 10\,s. To mitigate time-dependent systematic errors during spectral acquisition, a total of 64 individual scans each of \proton\ and \phos\ were taken in an alternating fashion automatically using a TopSpin script.  
\label{birdSubsection}

\subsection{Estimation of diastereomeric splitting}
\subsubsection{Titration with proton decoupling}
Peak-center frequency estimates were obtained from proton spectra (average of 32 transients) by fitting a sum of two absorptive Lorentzian doublet functions to \proton\ resonances from the methyl group of \probe\ around 1.54\,ppm. Diastereomeric splittings (\deltad ) in \proton\ spectra were determined by taking the difference of central frequencies between the fitted doublets. A similar procedure was repeated in \phos\ spectra (average of 64 transients) to determine diastereomeric splitting by fitting one Lorentzian doublet split by a frequency taken to represent \deltad. The amplitudes of the doublet peaks are allowed to vary independently to account for non-equal concentrations and formation rates of diastereomeric complexes \cite{Purity1968}. The fitting error of each \deltad\ measurement reported for the titration with proton decoupling is computed from the square root of the variance given by the fit (python scipy.optimize.curve\_fit) and the sample preparation error is estimated from 3 samples of identical \textit{S:R} composition ($\sigma_{^{1}\mathrm{H}}=1.34\%, \sigma_{^{31}\mathrm{P}}=2.12\%$). The error was included in the fitting of the comagnetometry plot shown in the supporting information. 

\subsubsection{Titration without proton decoupling and with BIRD}
Estimates of resonance frequencies were extracted from \proton\ data by fitting the sum of 2 transients with the sum of four complex Lorentzian functions a total of 32 times to include all of the 64 scans taken for each sample. Likewise, sums of 8 \phos\ spectra were fit using a sum of absorptive Lorentzian multiplets (without accounting for phase) a total of 8 times to fit all 64 spectra for each sample. The average \deltad\ of each sample was computed along with the standard deviation from the population of \deltad\ estimates. 

Uncertainty from sample preparation was determined by computing the standard deviation of \deltad\ estimates in both \proton\ and \phos\ spectra from 3 samples of identical \textit{S:R} composition, measured three times over 48\,h for a total of 9 measurements for each nucleus. The standard deviation was then divided by the sample mean to generate relative uncertainty. Total uncertainty was calculated by combining the standard deviation of frequencies given by fitting, $\sigma_\mathrm{fit}$, and the uncertainty due to sample preparation, $\delta_{s}$ at each point in quadrature as \[T.U. = \sqrt{\sigma_\mathrm{fit}^2 + (\Delta_d \delta_{s})^2}\] with $\delta_{s}^{^1\mathrm{H}}=1.34\%$ and $\delta_{s}^{^{31}\mathrm{P}}=2.12\%$ for all points. These values were then used to generate the final fitting estimates shown in figure \ref{fgr:comagPlot}.

\subsection{Linear regression of diastereomeric splitting}
The values for \deltad\ were plotted and fit with a linear model, $y_i = ax_{i} + b$ where $y_i$ is the extracted \phos\ diastereomeric splitting and $x_i$ is the extracted \proton\ diastereomeric splitting. Linear regression was accomplished using the Minuit package in python which was set to minimize $\chi^2$ as \[\chi^2 = \sum_i\left( \frac{y_{i}-y(x)}{T.E.} \right)^2,\] where $y(x)$ is the \phos\ splitting given by the model at $x_i$ and $T.E.$ is the total error associated with each point, \[T.E. = \sqrt{(a \sigma_{x_{i}})^2 + \sigma_{y_{i}}^2 + 2a\rho_{xy}\sigma_{x_{i}}\sigma_{y_{i}}}\,,\] with $\rho_{xy}$ calculated as Pearson's r. The error associated with the y-intercept $b$ in the final plot is the standard deviation computed from the covariance matrix given by the fit multiplied by 1.96 to reflect a 95\% confidence interval. A similar procedure was followed for the 3-dimensional measurement with two proton \deltad\ values extracted from \probe. Weighted averages ($\Bar{x}$) are computed using the formula 
\[\Bar{x}=\frac{\sum_i{x_i/\sigma^2_i}}{\sum_i{1/\sigma^2_i}}\] and associated error $\delta\Bar{x}^2$
\[\delta\Bar{x}^2=\frac{1}{\sum_i{1/\sigma^2_i}}.\]

\subsection{Quantum chemical calculations}
For an estimation of the expected PV splitting due to fundamental weak interactions in the studied phosphorous compounds we employed quasi-relativistic density functional theory (DFT) calculations. All these calculations were performed with a modified version
\cite{berger:2005,berger:2005a,isaev:2012,gaul:2020,bruck:2023,colombojofre:2022,zulch:2022} of a two-component program \cite{wullen:2010} based on Turbomole.\cite{ahlrichs:1989} We sampled the space of conformers of cyclohexyl \textit{P}-methylphosphonamidate with CREST \cite{pracht:2024}. The resulting 48 conformers were subsequently optimized at the level of non-relativistic restricted Kohn-Sham (RKS) calculations employing the hybrid exchange-correlation functional PBE0 \cite{perdew:1996,adamo:1999} with a triple-$\zeta$ Ahlrichs basis set (def2-TZVPP) \cite{weigend:2005} using the program package Turbomole 7.8 \cite{TURBOMOLE,balasubramani:2020}. For conformational averaging we computed vibrational frequencies and thermodynamic corrections to total energies with the freeh program of Turbomole for standard conditions (temperature $298.15\,\mathrm{K}$ and pressure $1\,\mathrm{hPa}$). All 48 conformers were found to be minima on the potential energy hypersurface. By comparison of energies, vibraional frequencies and molecular structures 23 unique conformers could be identified. The unique conformers of \textit{(R)}-cyclohexyl \textit{P}-methylphosphonamidate are provided in the supplementary material in form of coordinate files in xyz-format. Properties were averaged with Boltzmann weighing. In the supplementary material a table is provided, which contains all individual NMR properties. A single conformer of cyclohexyl \textit{N},\textit{N}-diethyl-\textit{P}-methylphosphonamidate was optimized at the same level of theory. Phosphoric acid H$_3$PO$_4$ was used as NMR standard to compute $^{31}$P chemical shifts. The molecular structure of H$_3$PO$_4$ was optimized at the same level of theory. Subsequently, quasi-relativistic densities were computed at the level of complex generalized Kohn-Sham (cGKS) within local density approximation (LDA) using the X$\alpha$ exchange
functional \cite{dirac:1930,slater:1951} and the VWN-5 correlation
functional \cite{vosko:1980} in a hybrid version with 50\,\% Fock exchange by Becke (BHandH)
\cite{becke:1993}. We employed an augmented all-electron correlated uncontracted Gaussian-type Dyall basis set (dyall.aae3z)\cite{dyall:2006} with additional sets of seven s-type and seven p-type functions with exponential factors composed as an even-tempered series $\zeta_i=\zeta_0/2^i$ with $\zeta_0=10^9\,a_0^{-2}$  (dyall.aae3z+sp) at $^{31}$P and the $^{13}$C and the protons at the methyl group bound to $^{31}$P as well as for N, O in cyclohexyl \textit{P}-methylphosphonamidate and H$_3$PO$_4$. For N and O (only in case of  cyclohexyl \textit{N},\textit{N}-diethyl-\textit{P}-methylphosphonamidate) and all other H and C atoms the IGLO-III basis set \cite{kutzelnigg:1991} was used as well as for O and N in cyclohexyl \textit{N},\textit{N}-diethyl-\textit{P}-methylphosphonamidate. Relativistic effects were considered at the level of two-component zeroth order regular approximation (ZORA) using the model potential approach by van W\"ullen to alleviate the gauge dependence of ZORA \cite{wullen:1998}. The model potential was applied with additional damping \cite{liu:2002}. Spectroscopic properties were computed with the toolbox approach of Ref.~\cite{gaul:2020} and response functions were computed as detailed in Refs.~\cite{bruck:2023,colombojofre:2022}. Conventional NMR shieldings were computed as described in Ref.~\cite{koziol:2024}. Indirect nuclear spin-spin couplings were computed as detailed in Ref.~\cite{colombojofre:2022} employing magnetogyric ratios $\gamma_\mathrm{^{1}H}=5.58569468\,\mu_\mathrm{N}$,  $\gamma_\mathrm{^{13}C}=1.4048236\,\mu_\mathrm{N}$, $\gamma_\mathrm{^{31}P}=2.2632\,\mu_\mathrm{N}$ as given in Ref.\,\cite{stone:2005}. In all calculations a common gauge origin of the homogeneous magnetic field  was employed. The gauge origin was placed at the respective atom, whose NMR chemical shielding was studied. PV frequency shifts to the isotropic NMR shielding of nucleus $A$ were computed in second order perturbation theory, below for convenience represented in a four-component sum-over-states formulation, using the following effective interaction Hamiltonians \cite{}:
\begin{multline}
\nu_\mathrm{PV} =\frac{-e\,c\,B_0\,\lambda_{\mathrm{PV}}(1-\sin^2\theta_\mathrm{W})G_\mathrm{F}}{h2\sqrt{2}}\\
\times\mathrm{Tr}\left[2\mathrm{Re}\sum\limits_{a \ne 0}\frac{\Braket{0|\sum\limits_{i=1}^{N_\mathrm{elec}}\vec{\boldsymbol{\alpha}}_i\rho_A(\vec{r}_i)|a}\Braket{a|\sum\limits_{i=1}^{N_\mathrm{elec}}[\vec{r}_{iA}\times\vec{\boldsymbol{\alpha}}_i]^\mathrm{T}|0}}{E_0-E_a}\right]/3,
\end{multline}
Here $\Braket{a|\hat{A}|b}$ denotes matrix elements of a given operator $\hat{A}$ between two many-electron
wave functions and $\Ket{0}$, $\Ket{a}$ denote wave functions of a 
ground-state reference and excited electronic states with energies $E_0$ and $E_a$
respectively. $\mathrm{Tr}[\bm{A}]$ is the trace of matrix $\bm{A}$ $c$ is the speed of light, $h$ is the Planck constant, $B_0$ is external homogeneous magnetic field of strength, $\varrho_A$ is the normalized nuclear density distribution, $\vec{r}_{ab}=\vec{r}_a-\vec{r}_b$ is the relative position of two particles and $\vec{v}^\mathrm{T}$ refers to the transpose of a vector $\vec{v}$. In calculations of PV NMR shieldings Fermi's weak coupling constant $G_\mathrm{F}=2.22249\times10^{-14}\,E_\mathrm{h}a_0^3$, $\mathrm{sin}^2\theta_\mathrm{W}=0.2319$ with $\theta_\mathrm{W}$ being the Weinberg angle and a nucleus dependent coupling strength parameter of $\lambda_\mathrm{PV}=-1$ for all nuclei were employed in order to be consistent with previous studies on PV contributions to NMR shielding constants in chiral molecules. We emphasize here that our reported PV NMR parameters are effective in the sense that they have to be scaled finally by corresponding nuclear-structure dependent terms that account most importantly also for the nuclear anapole moments of the specific isotope. The Dirac matrix $\vec{\boldsymbol{\alpha}}$ is defined as 
$\vec{\boldsymbol{\alpha}}=\begin{pmatrix}\boldsymbol{0}&\vec{\boldsymbol{\sigma}}\\\vec{\boldsymbol{\sigma}}&\boldsymbol{0}\end{pmatrix}$ where $\vec{\boldsymbol{\sigma}}$ is the vector of Pauli matrices.
For a detailed derivation of PV NMR shieldings within ZORA see Ref.~\cite{nahrwold:09}. The Coulomb potential of the nuclei was modeled in all calculations assuming a finite spherical Gaussian-shaped nuclear charge density distribution $e Z_A \varrho_A \left( \vec{r} \right) = e Z_A \frac{\zeta_A^{3/2}}{\pi ^{3/2}} \text{e}^{-\zeta_A \left| \vec{r} -
\vec{r}_A \right| ^2}$ with $\zeta_A = \frac{3}{2 r^2
_\text{nuc},A}$ and the root-mean-square radius $r_{\text{nuc},A}$ was chosen as suggested by Visscher and Dyall \cite{visscher:1997}, where nuclear mass numbers where chosen as nearest integers to the natural mass of the element. Nuclear magnetization distributions were assumed to be point-like in all calculations. We define PV splitting as $\Delta_\mathrm{PV}=\nu_\mathrm{PV}(R)-\nu_\mathrm{PV}(S)$, where $\nu_\mathrm{PV}(R)$, $\nu_\mathrm{PV}(S)$ are the PV NMR frequency shifts for the $(R)$ and $(S)$ enantiomer respectively.

\section{Results and Discussion}

\subsection{Estimation of expected PV splittings}
PV splittings of $^{31}$P NMR signals between two enantiomers where computed as detailed in the previous section to be $\Delta_\mathrm{PV}(^{31}\mathrm{P}) = -0.7\,\mathrm{\upmu Hz}$ for cyclohexyl \textit{P}-methylphosphonamidate (conformationally averaged) and $\Delta_\mathrm{PV}(^{31}\mathrm{P}) = 0.2\,\mathrm{\upmu Hz}$ for cyclohexyl \textit{N},\textit{N}-diethyl-\textit{P}-methylphosphonamidate (single conformer) when assuming an external homogeneous magnetic field of strength 20~T. For comparison with experiment we computed with the same methodology the conventional $^{31}$P-NMR chemical shifts relative to H$_3$PO$_4$ (computed isotropic shielding constant is $\sigma=319.3\,\mathrm{ppm}$) for those two compounds. The $^{31}$P-NMR chemical shifts were found to be 24 ppm for cyclohexyl \textit{P}-methylphosphonamidate (conformationally averaged) and 33 ppm for cyclohexyl \textit{N},\textit{N}-diethyl-\textit{P}-methylphosphonamidate (single conformer). These computed chemical shifts deviate considerably from the experimental chemical shifts by 25\,\% and 9\,\%, respectively. The methyl-$^1$H-$^{31}$P ${}^2J$-coupling in cyclohexyl \textit{P}-methylphosphonamidate (conformationally averaged) was computed to be $-12$ Hz, which deviates 28\,\% from the experimental absolute value of 16.8 Hz shown in figure \ref{fgr:fittingPlot}, a similar deviation as for the chemical shift. Our calculation suggests a negative sign of the $^1$H-$^{31}$P ${}^2J$-coupling. 

\begin{figure}
\centering
  \includegraphics[width=3.3in]{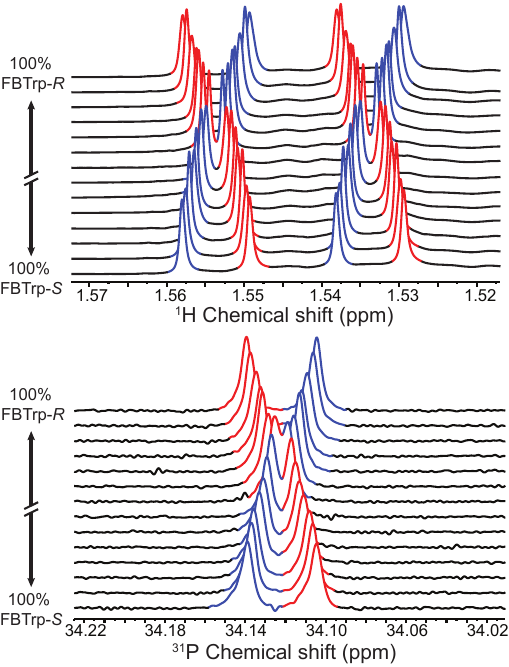}
  \caption{Titration of enantiomeric ratio of \fmoc\ with resultant shift in methyl-\proton\ (top) and \phos\ (bottom) NMR resonances of \probe.  \textcolor{blue}{Blue} and \textcolor{red}{red} highlighted peaks indicate signals arising from complimentary \textit{S,S/R,R} and opposite \textit{R,S/S,R} diastereomeric enantiomers, with the frequency depending on the net chirality of the chiral solvating agent. A smooth transition to higher or lower frequencies at intermediate \fmoc\ enantiomeric ratios occurs because the system is undergoing rapid exchange. Phosphorus spectra are shown with \proton\ decoupling for clarity, and values of 56\% \textit{S} to 44\% \textit{S} are omitted due to difficulty extracting peak frequency estimates from spectral overlap.}
  \label{fgr:stackedTitrationSpectra}
\end{figure}

Here, we want to emphasize that computed chemical shifts for the two compounds are small compared to the typical range of chemical shifts in $^{31}$P NMR spectroscopy, which may be one of the reason why an opposite trend is seen in comparison to the experimental chemical shifts of the two compounds. Moreover, the use of a single conformer for cyclohexyl \textit{N},\textit{N}-diethyl-\textit{P}-methylphosphonamidate may be an important source of error. For cyclohexyl \textit{P}-methylphosphonamidate we observed differences of up to 10 ppm ($\sim40\,\%$) between different conformers. Other likely error sources are the employed exchange-correlation functional BHandH, which is not explicitly designed for calculations of NMR parameters, and the molecular structures, which were computed at the DFT level as well. Moreover, errors of computed chemical shifts can be partially attributed to the basis set used in this work, which was not large enough to completely suppress the gauge origin dependence. For example, a calculation with a gauge origin shifted by $10~a_0$ in every spatial direction increases chemical shifts by 3\,ppm, i.e. a change of 10\,\%. Here, we like to emphasize that the dependence on the gauge origin is negligible for PV NMR shifts, which was found to be below 1\,\%, in agreement with previous PV-NMR calculations.\cite{nahrwold:09,nahrwold:2014} Further uncertainties of the calculated chemical shifts and J coupling constants could be due to solvent effects, which were neglected in our calculations, wherein the molecules are in vacuum. For the present purpose, i.e. estimating the expected size of PV splittings, however, we consider our calculations to be sufficiently accurate.

The PV effects are predicted to be two to three orders of magnitude below the expected experimental resolution and therefore are not detectable in the present measurements as assumed before. In molecules composed of light or medium heavy elements only, spin-orbit coupling effects are typically small and PV NMR shielding tensors are then expected to scale with the nuclear charge number $Z$ as about $Z^2$.\cite{Gorshkov1982,laubender:2003,nahrwold:09} Therefore, PV effects on the internal comagnetometer signal from the $^1$H nucleus should be at least two orders of magnitude lower than PV splittings on the $^{31}$P nucleus, which is confirmed by our numerical calculations: The PV splitting for $^1$H located at the methyl group in cyclohexyl \textit{P}-methylphosphonamidate is found to be $\Delta_\mathrm{PV}(^{1}\mathrm{H}) \lesssim5 \times10^{-4}\,\mathrm{\upmu Hz}$.

\subsection{Resolution of diastereomeric phosphorus complex}
Li and Raushel published a method for resolving chiral oxophosphoranes using substituted amino acids such as tryptophan \cite{Li2007}. We adopted a pair of diastereomer-forming compounds based on their work, namely a chiral phosphonamidate, cyclohexyl \textit{P}-methylphosphonamidate (\probe), and N-fluorenylmethyloxycarbonyl-N’-tert-butyloxycarbonyl-tryptophan (\fmoc), pictured in figure\,\ref{fgr:figure1}. The \probe\ is referred to as a sensor molecule because it contains the high-$Z$ nucleus (in this example, \phos) which should be sensitive to P-odd effects caused by the weak interaction. Though, for \phos, any PV effects are expected to be below the detection limit, the target of this study is to determine systematic errors present in this approach and develop strategies to mitigate them. 

The target resonances both originate from \probe, namely the \phos\ resonance of the chiral phosphorus center and the \proton\ resonances of the adjacent methyl group. Preliminary tests using chloroform-d as a solvent show that an equimolar combination of \probe\ and \fmoc\ produces the largest overall diastereomeric splitting (\deltad) between enantiomers in both \phos\ and \proton\ spectra. Several other solvents were tested with the aim of generating the greatest \deltad\ in both \phos\ and \proton\ NMR signals of \probe. \textit{ortho}-Dichlorobenzene-d$_4$ and dichloromethane-d$_2$ also produced considerable \deltad\ in both \proton\ and \phos\ spectra. However samples prepared in chloroform-d showed the greatest overall \deltad\ in both \proton\ (6.5 Hz) and \phos\ (13.3 Hz) spectra (Table 1). 

\begin{table}[h]
\small
\caption{Diastereomeric splitting of 10\,mM \probe\ with 10\,mM \fmocS\ in fully deuterated organic solvents at 20\,T, 298\,K}
\label{tbl:example1}
\begin{tabular*}{0.48\textwidth}{@{\extracolsep{\fill}}lll}
    \hline
    Solvent & \deltad\proton\ (Hz) & \deltad\phos\ (Hz)\\
    \hline
    Acetone-d$_6$   & 2.1    & 4.9\\
    Tetrahydrofuran-d$_8$ & 0 & 9.5\\
     \textit{ortho}-Dichlorobenzene-d$_4$ & 4.1 & 12.2\\
    Chloroform-d & 6.5 & 13.3 \\
    Dichloromethane-d$_2$ & 3.1 & 17.5\\
    Dimethyl sulfoxide-d$_6$ & 0 & 0\\
    \hline
  \end{tabular*}
  \caption*{\deltad\proton\ denotes diastereomeric splitting values for the methyl-\proton\ resonance of \probe.}
\end{table}

It was also noted that at higher concentrations of \fmoc, with \probe\ concentrations held at 10\,mM, \phos\ spectral lines exhibited a nonlinear shift towards higher frequencies, while a similar shift was seen in \proton\ spectra of the \probe\ methyl group towards lower frequencies under the same conditions (see SI). This is expected to occur due to  rapid chemical exchange between \probe\ and \fmoc. 

A second, chemically similar phosphonamidate -- cyclohexyl N,N-diethyl-\textit{P}-methylphosphonamidate (N,N-\probe) -- was also examined in solution with \fmoc\ at several ratios of \fmoc\ to N,N-\probe\ in chloroform-d. While modest \deltad\ was observed in \proton\ spectra, none of the tested conditions resulted in reproducible \deltad\ in \phos\ spectra (see SI). We therefore did not include this molecule in further experiments. Diminished diastereomeric splitting indicates that the association of \probe\ and \fmoc\ is reduced by the presence of ethyl groups, possibly due to steric effects. It is also likely that the amino group of \probe\ participates in hydrogen-bonding with \fmoc\ in addition to the phosphoryl oxygen atom previously proposed as the predominant binding site \cite{Li2007}. This additional interaction could lead to several different H-bonding conformations, one of which is depicted in figure \ref{fgr:figure1}, with additional possibilities shown in the supporting information. The additional contact afforded by a second H-bonding site may be responsible for the relatively strong diastereomeric splitting observed in \probe\ as opposed to other phosphorus-containing molecules examined by Li and Raushel. 

Having chosen the concentrations, we proceeded with collection of high-field NMR spectra. The enantiomeric ratio of \fmoc\ was titrated from 100\% \fmocS\ to 100\% \fmocR\ in samples containing 10\,mM \probe\ and 10\,mM \fmoc\ (total), the result of which can be seen in spectral form in figure\,\ref{fgr:stackedTitrationSpectra}. Frequency estimates were extracted by fitting analytical Lorentzian functions to the Fourier transformed time domain signal originating from methyl-\,\proton\ and \phos\ spins of \probe\ (see figure\,\ref{fgr:fittingPlot}). Resultant \deltad\ values extracted from the data, taken as the difference between the frequency of signals from each diastereomeric pair (\textit{S,S}/\textit{R,R} and \textit{S,R}/\textit{R,S}), are plotted against each other (see supplementary information). These values were then fit using linear-regression accounting for error in both the \proton\ and \phos\ axes, to obtain an estimate of the residual \phos\ \deltad\ at the racemic point ($\Delta_d{^{31}\mathrm{P}}(0)$).  

In total, two such enantiodiscriminatory titrations were performed from which $\Delta_d{^{31}\mathrm{P}}(0)$ could be extracted: one with \proton\ spectra collected using a simple 90\textdegree\ pulse and acquire sequence and \phos\ spectra collected with a 90\textdegree\ pulse and \proton\ decoupling; and another titration where \proton\ spectra were collected using a BIRD pulse sequence and \phos\ spectra with no \proton\ decoupling.

\begin{figure}
\centering
    \includegraphics[width=3.3in]{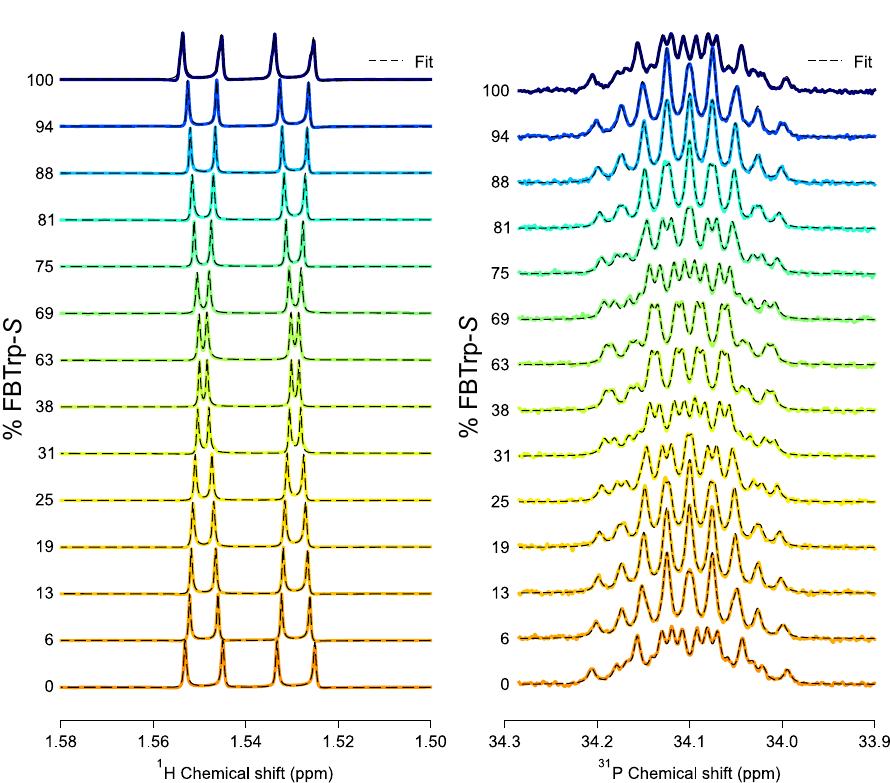}
    \caption{\proton\ (left) and \phos\ (right) NMR spectra at 850 MHz from which \deltad\ values were extracted for the final comagnetometry plot shown in figure\,\ref{fgr:comagPlot}. \proton\ spectra were collecting using a BIRD pulse sequence (see supporting information) and fit with the sum of four complex Lorentzian lines with independent phase and peak center. \phos\ spectra were acquired using a simple 90\textdegree pulse sequence without \proton\ decoupling and fit with a sum of two multiplets (doublet of quartets, AXY\textsubscript{3} spin system) with no phase, assuming purely absorptive lineshapes. Experimentally obtained spectra are overlaid with the best fit line at each sample composition.}
    \label{fgr:titration3}
\end{figure}

In the first experiment, two \proton\ resonances and one \phos\ resonance were analyzed to extract \deltad. From these, residual \phos\ splitting values of $\Delta_d{^{31}\mathrm{P}}(0)=-170\pm100$ mHz, and $\Delta_d{^{31}\mathrm{P}}(0)=-200\pm120$ mHz (weighted average, $\Bar{\Delta}_d{^{31}\mathrm{P}}(0)=-190\pm80$) were extracted using 2-dimensional linear regression. Comparing all three gave $\Delta_d{^{31}\mathrm{P}}(0,0)=-190\pm120$ mHz following 3-dimensional linear regression. Comparing \deltad\ of both \proton\ multiplets yields $\Delta_d{^1\mathrm{H}}(0) = 3 \pm 36$ mHz which indicates the \proton\ measurement does not contain systematic error or other physical effects that lead to a non-zero racemic point intercept. Once the \phos\ measurement is factored in however, there clearly arises a shift away from zero, which is unexpected.

The BIRD sequence in the second titration was calibrated according to the $J$-coupling between the methyl-\proton s and the chiral \phos\ to suppress broad peaks around the \proton\ mutiplet. This had the side effect of suppressing the second proton multiplet used in the \proton-\proton\ linear regression and 3-dimensional analysis of first titration. Fortunately, the precision of this 2-dimensional measurement is higher than both of those in the first titration by a factor of approximately 2, yielding $\Delta_d{^{31}\mathrm{P}}(0)=-56\pm61$ mHz. Additionally, because each measurement was taken in the form of 64 individual scans for each nucleus rather than as averages, as in the first, it was possible to characterize the uncertainty associated with this measurement using conventional statistical considerations in addition to computing the uncertainty from the covariance matrix given by the fit. Thus, the sample mean and standard deviation of the extracted frequency estimates was used to generate the fit shown in figure\,\ref{fgr:comagPlot} along with error from fitting and sample preparation.

In fitting the data displayed in figure \ref{fgr:comagPlot}, a minimum $\chi^2$ of 4.2 is reached, accounting for error from sample preparation and from fitting spectra in both \proton\ and \phos\ measurements (see experimental section). Since $\chi^2=4.2$ is less than the degrees of freedom in our measurement ($dof=12$), this indicates that either the errors are over-estimated or there are strong correlations between the measurements \cite{Schmelling1995}. Thus the errors in both \proton\ and \phos\ are scaled by a factor ($\sqrt{\chi^2/dof}=0.58$) such that $\chi^2$ is equal to the degrees of freedom. 

\subsection{Discussion of possible systematic error}

\subsubsection{Nonlinearity of \phos\ diastereomeric splitting}
Ideally, we would want to work in a regime where the diastereomeric splitting of both protons and the heavy nuclei are linear in both the enantiomeric ratio and the concentration of the chiral solvating agent. Unfortunately, the latter is far from being satisfied in the present case as can be seen in both \proton\ and \phos\ spectra (see SI). As shown below, despite the nonlinear concentration dependence at the chosen 1:1 ratio, our ``comagnetometry'' approach is still able to provide enhanced resolution compared to the NMR linewidth. Since nonlinear concentration dependence is a likely source of systematic error, it would be best to find a system free from this effect or operate in a linear range for future experiments with heavier nuclei.

Comparing proton-proton and proton-phosphorus \deltad\ helps determine in which measurement systematic error arises. Since the proton-proton y-intercept measurement is consistent with zero ($\Delta_d{^1\mathrm{H}}(0)=3\pm 36$ mHz), while the proton-phosphorus measurement is not for both the methyl-\proton\ and cyclohexyl-\proton\ resoncances (see SI), this is a good indication that the systematic error lies in the measurement of the \phos\ nucleus. As previously mentioned, nonlinear changes of \deltad\ in one nucleus with respect to CSA enantiomeric ratio are uncompensated by \deltad\ in the second nucleus and can contribute to systematic error in the measurement of $\Delta_\mathrm{PV}$.

\begin{figure}[h]
\centering
  \includegraphics[width=3.3in]{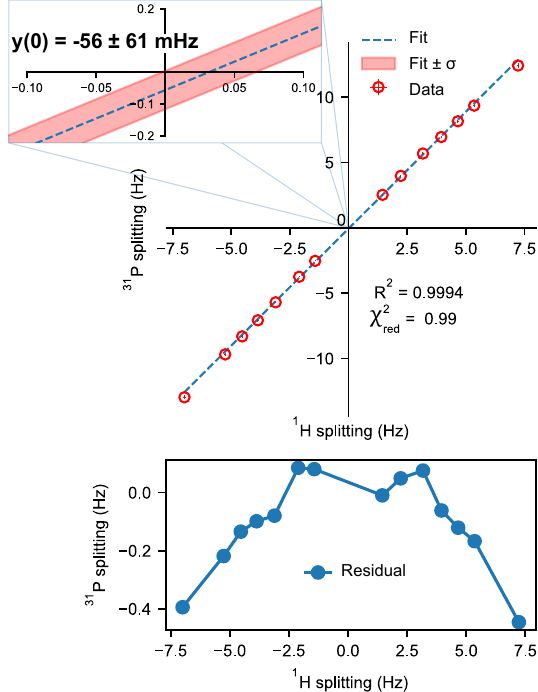}
  \caption{Comagnetometry plot showing diastereomeric splitting (\deltad) of \phos\ spectra as a function of \deltad\proton. Each point represents the average \deltad\ $\pm$ uncertainty from 64 scans in a 20.0\,T NMR spectrometer at 298\,K. The region around the origin in enlarged to show the y-axis (i.e. high-$Z$ axis) intercept of the linear fit. A reduced chi squared ($\chi_\mathrm{red}^{2}$) value is used to calibrate the error estimates on each point such that $\chi_\mathrm{red}^{2}=1$. The residual plot below indicates there is some deviation from linearity.}
  \label{fgr:comagPlot}
\end{figure}

A critical assumption in our approach is that systematic errors arising from errors in sample preparation are largely removed by nuclear co-sensing, as changes in \deltad\ due to most sources of error -- sample preparation, temperature drifts, viscosity, etc. -- in \phos\ should be compensated by \proton\ \deltad. To test this assumption, three samples were prepared with the same stock solutions, implements, measured with the same pulse sequence, and processed to extract \deltad. As shown in the supplementary information, we observed a nonlinear dependence of \phos\ and \proton\ \deltad\ across the three samples. Strikingly, the \deltad\ measured over the course of 48 hrs showed less variablity compared to measurements between samples. This variability has been incorporated into the final measurement as relative uncertainty ($\delta_{^1\mathrm{H}} = 1.54\%$, $\delta_{^{31}\mathrm{P}} = 2.14\%$) for each point shown in figure \ref{fgr:comagPlot}. It is noted that this has the effect of exaggerating the error in points distal to the origin, creating a bias towards points nearer the origin \cite{Schmelling1995}. 

The \phos -\proton\ $J$-coupling network was characterized using the ANATOLIA software package \cite{Cheshkov2018}. For this, the \phos\ multiplet structure of \probe\ in chloroform-d was modeled as an AXY$_3$ spin system model (see SI). This generated coupling constants of ${}^{2}J_\mathrm{PCH_3}$ = 16.8 Hz and ${}^{3}J_\mathrm{PH}$ = 9.1 Hz for the \phos\ $J$-coupling to the methyl protons and to the cyclohexyl proton nearest the phosphorus center, respectively (see SI). This assignment is further supported by \proton\ spectra of both \probe\ and its N,N-diethyl analog which both show a \proton\ multiplet with an identical splitting pattern at similar chemical shifts. An attempt to model the spin system as an AX$_2$Y$_3$ system including coupling to the amino protons of \phos\ was made; however the simulated spectra produced using ANATOLIA were unable to match the experimentally observed spectra which indicates these protons are participating in rapid chemical exchange.

\subsubsection{Systematic error from instrumentation}
Another proposed source of systematic error in the first enantiodiscriminatory titration is the inverse gated proton decoupling pulse applied during the acquisition of \phos\ spectra. In principle, imperfections in the pulse coil array could produce a small magnetic field component parallel to the axis of $B_0$, modulating \deltad\ in \phos\ spectra that would not be compensated by \proton\ measurements since the same field is not applied during \proton\ acquisition. For example, if a decoupling field of 1\,mT is applied, and 10\% of this leaks into $B_0$, that would constitute an additional field of 100\,$\upmu$T. For a \phos\ splitting of 12\,Hz at 20 T (0.1\,ppm), this field would produce an additional splitting of 60\,$\upmu$Hz, which is in the range of a problematic systematic error. To avoid this possibility, the second titration was completed without the use of proton decoupling in \phos\ spectra. However, removing \proton-decoupling had a complicating effect on the analysis of \deltad\ due to the appearance of additional peaks caused by $J$-coupling to methyl and cyclohexyl protons (see figure \ref{fgr:fittingPlot} (d)) and a reduction in signal-to-noise ratio. For ease of interpretation, the \phos\ spectra collected using \proton-decoupling are shown in figure\,\ref{fgr:stackedTitrationSpectra}, but the reader should note that the \phos\ spectra used to generate the final comagnetometer plot seen in figure\,\ref{fgr:comagPlot} are shown in figure\,\ref{fgr:titration3}.

Drifts in chemical shift caused by changes in temperature are also possible over the course of a measurement with signal averaging. Though largely mitigated by modern instrumentation, the effects of this in our experiments was assessed in the second titration by taking 64 individual scans of \proton\ and \phos, alternating between \proton\ and \phos, for each scan. This allowed the spectra of each nucleus to be processed individually, rather than as a sum as is usual in NMR experiments, allowing statistical treatment of \deltad\ estimates. Time course frequency estimates show that in some samples there are definite drifts in the measurements of \deltad\ over time (see SI) despite using deuterium-locking and a temperature controlled probe.

\subsubsection{Uncertainty increased by overlapping resonances}
\begin{figure}
\centering
  \includegraphics[width=3.3in]{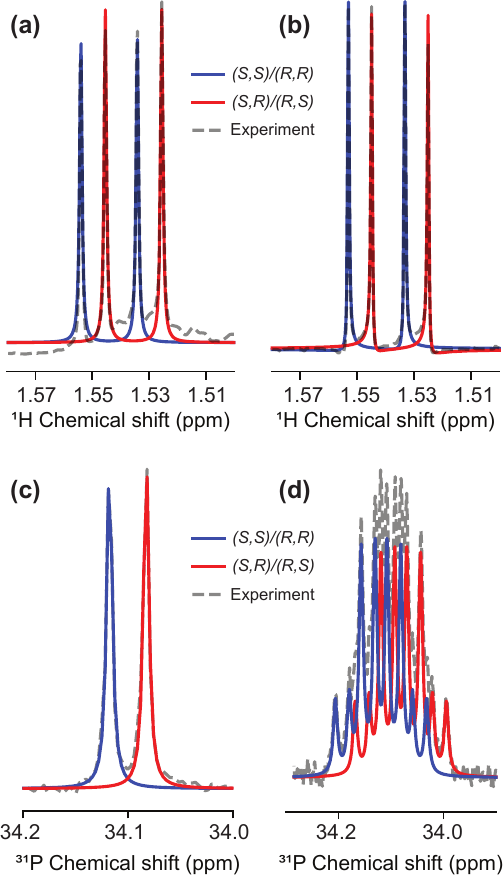}
  \caption{Plots illustrating the fitting procedure used to extract \deltad\ from \proton\ and \phos\ NMR spectra, with pairs of diastereomeric enantiomers (\textit{S,S/R,R} and \textit{S,R/R,S}) highlighted in \textcolor{blue}{blue} or \textcolor{red}{red}. (\textbf{a}) \proton\ spectrum of a racemic mixture of \probe\ and 100\% \fmocS\ representing maximal diastereomeric splitting. Broad resonances from protons of the cyclohexyl group overlap with the methyl resonances used for fitting, reducing the goodness of fit and, subsequently, the precision of frequency estimates. (\textbf{b}) The same spectral region after utilizing a BIRD pulse sequence to remove broad peaks surrounding the methyl resonances. This sequence uses pulses to exploit methyl-\proton-\phos\ $J$-coupling of 16.8\,Hz to keep the magnetization of coupled protons oriented along $B_z$ while rotating all other magnetization by 180 degrees. A brief relaxation delay of 100\,ms allows negative magnetization to relax to zero before application of a 90\textdegree pulse for acquisition. (\textbf{c}) \phos\ spectrum of racemic \probe\ and 100\% \fmocS\ with with inverse gated proton decoupling applied during acquisition. (\textbf{d}) \phos\ spectrum of the same sample without proton decoupling. The signal is a sum of multiplets shown in blue and red corresponding to pairs of epimers.}
  \label{fgr:fittingPlot}
\end{figure}

Upon initial inspection, the spectral lines of methyl protons of \probe\ exhibited extensive overlap with broad resonances from the cyclohexyl group. This introduced a positive offset when fitting the baseline of the spectra shown in figure\,\ref{fgr:fittingPlot} (a), causing a loss in precision of \deltad\ estimates. To mitigate this, a second titration was performed with a BIRD pulse sequence (see section\,\ref{birdSubsection} and figure\,\ref{fgr:fittingPlot} (b)) to suppress non-\phos\ coupled spins. Removing these baseline distortions increased the precision of frequency estimation by a factor of 5. However, using a BIRD sequence introduced two side effects: 1) phase distortions in the methyl-\proton\ peaks, and 2) suppression of the second \proton\ multiplet, removing the ability to extract a second \deltad\proton\ measurement.The first effect is easily seen comparing \proton\ spectra in figure\,\ref{fgr:fittingPlot}, as the two Lorentzian doublets used in the fit seem to indicate that the peaks of each set of diastereomeric enantiomers have acquired an equal and opposite phase component. 

\subsubsection{Multi-dimensional diastereomeric splitting titration}
Another avenue to deal with persistent systematic errors is measurement of additional nuclei which display diastereomeric splitting within the same molecule. Doing so may allow the elimination of additional dimensions of systematic errors and an increase in precision. 

To this end, a second peak displaying diastereomeric splitting, namely the cyclohexyl proton closest to the \phos\ center, was analyzed using a similar fitting procedure as the other peaks reported in this study (see SI). This \deltad\ was then correlated with the two other \deltad\ measurements to generate a 3-dimensional plot which was fit to generate an estimate of residual PV shift in \phos\ spectra (see SI), giving a $z$-intercept estimate (where PV effects are expected to appear in high-$Z$ nuclei) of $-190 \pm 120$ mHz. This shows that the precision in this measurement with this system is not capable of resolving the predicted mHz PV energy shifts in heavy nuclei of chiral molecules. Furthermore, since the splitting is expected to be several $\upmu$Hz in \phos, there is still a significant systematic error present in this measurement since zero is not included. 

This along observations of nonlinear changes in \phos\ \deltad\ indicate that our comagnetometry approach is influenced by more than contamination of the system by additional chiral molecules. This implies that concentrations must be very carefully controlled to reach the required mHz precision needed to observe PV in NMR.

\subsection{Other considerations}
Tautomerization is believed to produce the phosphine imide as well as the dominant amine form of \probe. This phosphine imide is expected to be highly reactive \cite{Zhu2021, Niecke1996, Niecke2001, Niecke2006} and could lead to the creation of species seen in samples at higher concentrations of \phos\ and \fmoc\ shown and also in the supplementary information. 

Distributions of cyclohexane ligand conformations could lead to broadening of proton or phosphorus signals through Weak (>1\,Hz) multi-bond $J$-couplings. This would contribute to broadening of spectral lines, though is not expected to produce an asymmetric shift away from 0 Hz at the racemic point. 

Diastereomeric splitting is caused by groups in the CSA which induce changes in electronic charge distribution in the target molecule, and may also be enhanced by the chiral-induced spin selectivity (CISS) effect \cite{Naaman2020, Kapon2023}. The CISS effect leads to preferential transfer of electrons through chiral molecules based on polarization state, which when coupled to nuclei may cause shifts in frequency. This contribution is expected to be symmetric with respect to chirality inversion and not contribute to residual splitting at the racemic point. CISS may lead to different $T_1$ times for \textit{S} and \textit{R} enantiomers, however the extent of this has not been quantified \cite{Georgiou2024}.

\section{Conclusions}
The goal of this study was to asses sources of error in determining frequency estimates from NMR experiments using diastereomeric complexes containing an intermediate-$Z$ atom (\phos). Minimizing sources of error is crucial to observing parity violating contributions of the weak interaction to the chemical shift tensor of molecules, which is expected to be of the order of mHz for high-$Z$ nuclei. 

We show that using a CSA with chiral probe allows tunable diastereomeric splitting in two nuclei within the same complex, and find no ``show-stoppers'' for experiments with high-$Z$ systems, although effects of concentration dependence need to be carefully considered. The next steps would depend on finding/synthesizing an appropriate high-$Z$ system (perhaps, containing $^{203,205}$Tl, $^{207}$Pb,...).

\section*{Author Contributions}
DB, DAB, JE, EVD, KS, MW, and RG designed the experiments. RB proposed the initial compounds. AEW, KG and RB provided theoretical estimates of parity violating splitting in this system. DK synthesized the phosphorus probe molecule and carried out initial experiments with diastereomeric splitting. EVD prepared the samples, performed the experiments, and analyzed the data presented in the main text. EVD, DB, DK, and KG wrote the manuscript. All authors edited the manuscript.

\section*{Conflicts of interest}
There are no conflicts to declare.

\section*{Acknowledgements}
We thank Tanja Weil for support and Louis Bouchard, Yossi Paltiel, Gary Centers, Mohamed Sabba and Malcolm Levitt for useful discussions. This research was supported in part by the DFG Project ID 390831469: EXC 2118 (PRISMA+ Cluster of Excellence). We acknowledge the financial support by the Alexander von Humboldt Foundation in the framework of the Sofja Kovalevskaja Award. Financial support by the Deutsche Forschungsgemeinschaft (DFG, German Research Foundation) from the Collaborative Research Center “Extreme light for sensing and driving molecular chirality” (Projektnummer 328961117, CRC 1319 ELCH) is gratefully acknowledged. K.G. is indebted to the Funds of chemical industry (FCI) for funding. This work is generously supported by the by the Carl-Zeiss-Stiftung (HYMMS project P2022-03-044).



\balance


\bibliography{rsc} 
\bibliographystyle{rsc} 

\end{document}


\pagestyle{fancy}
\thispagestyle{plain}
\fancypagestyle{plain}{
\renewcommand{\headrulewidth}{0pt}
}

\makeFNbottom
\makeatletter
\renewcommand\LARGE{\@setfontsize\LARGE{15pt}{17}}
\renewcommand\Large{\@setfontsize\Large{12pt}{14}}
\renewcommand\large{\@setfontsize\large{10pt}{12}}
\renewcommand\footnotesize{\@setfontsize\footnotesize{7pt}{10}}
\makeatother

\renewcommand{\thefootnote}{\fnsymbol{footnote}}
\renewcommand\footnoterule{\vspace*{1pt}%
\color{cream}\hrule width 3.5in height 0.4pt \color{black}\vspace*{5pt}} 
\setcounter{secnumdepth}{5}

\makeatletter 
\renewcommand\@biblabel[1]{#1}            
\renewcommand\@makefntext[1]%
{\noindent\makebox[0pt][r]{\@thefnmark\,}#1}
\makeatother 
\renewcommand{\figurename}{\small{Fig.}~}
\sectionfont{\sffamily\Large}
\subsectionfont{\normalsize}
\subsubsectionfont{\bf}
\setstretch{1.125} 
\setlength{\skip\footins}{0.8cm}
\setlength{\footnotesep}{0.25cm}
\setlength{\jot}{10pt}
\titlespacing*{\section}{0pt}{4pt}{4pt}
\titlespacing*{\subsection}{0pt}{15pt}{1pt}

\fancyhead{}
\renewcommand{\headrulewidth}{0pt} 
\renewcommand{\footrulewidth}{0pt}
\setlength{\arrayrulewidth}{1pt}
\setlength{\columnsep}{6.5mm}
\setlength\bibsep{1pt}

\makeatletter 
\newlength{\figrulesep} 
\setlength{\figrulesep}{0.5\textfloatsep} 

\newcommand{\topfigrule}{\vspace*{-1pt}%
\noindent{\color{cream}\rule[-\figrulesep]{\columnwidth}{1.5pt}} }

\newcommand{\botfigrule}{\vspace*{-2pt}%
\noindent{\color{cream}\rule[\figrulesep]{\columnwidth}{1.5pt}} }

\newcommand{\dblfigrule}{\vspace*{-1pt}%
\noindent{\color{cream}\rule[-\figrulesep]{\textwidth}{1.5pt}} }

\makeatother

\onecolumn

\vspace{1em}
\sffamily

\noindent\LARGE
{\textbf{Supplementary information for: Towards detection of molecular parity violation via chiral co-sensing: the $^1$H/$^{31}$P model system}
}

\vspace{0.3cm} 

\noindent\large{Erik Van Dyke$^{\ast}$\textit{$^{a,b}$}, James Eills\textit{$^{c}$}, Kirill Sheberstov\textit{$^e$}, John Blanchard\textit{$^e$}, Manfred Wagner\textit{$^{f}$}, Robert Graf\textit{$^{f}$}, Andr{\'e}s Emilio Wedenig\textit{$^{g}$}, Konstantin Gaul\textit{$^{b,g}$}, Robert Berger\textit{$^{g}$}, Rudolf Pietschnig\textit{$^{h}$},  Denis Kargin\textit{$^{h}$},
 Danila A. Barskiy$^{\ast}$\textit{$^{a,b}$}, and 
 Dmitry Budker\textit{$^{a,b,i}$}} \\


\renewcommand*\rmdefault{bch}\normalfont\upshape
\rmfamily
\section*{}
\vspace{-1cm}


\nnfootnote{\textit{$^{a}$~Institute for Physics, Johannes Gutenberg University Mainz, 55128 Mainz, Germany.}}
\nnfootnote{\textit{$^{b}$~Helmholtz Institute Mainz, 55128 Mainz, Germany; GSI Helmholtz Center for Heavy Ion Research, Darmstadt, Germany.}}
\nnfootnote{\textit{$^{c}$~Institute of Biological Information Processing (IBI-7), Forschungszentrum Jülich, Jülich 52425, Germany}}
\nnfootnote{\textit{$^{d}$~Quantum Technology Center, University of Maryland, College Park, Maryland, MD, 20742 USA}}
\nnfootnote{\textit{$^{e}$~Laboratoire des biomolécules, LBM, Département de chimie, École normale supérieure, PSL University, Sorbonne Université, CNRS, 75005 Paris, France.}}
\nnfootnote{\textit{$^{f}$~Max Planck Institute for Polymer Research, 55128 Mainz, Germany.}}
\nnfootnote{\textit{$^{g}$~Institute for Chemistry, Philipps University Marburg, 35032 Marburg, Germany.}}
\nnfootnote{\textit{$^{h}$~Institute for Chemistry, University of Kassel, 34132 Kassel, Germany.}}
\nnfootnote{\textit{$^{i}$~Department of Physics, University of California at Berkeley, Berkeley CA 94720, USA.}}

\nnfootnote{\dag~Electronic Supplementary Information (ESI) available: [details of any supplementary information available should be included here]. See DOI: 10.1039/cXCP00000x/}

\nnfootnote{* Corresponding author.}



\section{NMR characterization of chiral phosphonamidate}
All experiments, if not stated otherwise, were carried out under exclusion of moisture and air under an inert argon atmosphere. Starting materials were purified and stored under argon. Methylphosphonic dichloride was prepared according to literature procedures. Starting from trimethylphosphite and catalytic amounts of methyl iodide, dimethyl methylphosphonate was prepared in an Arbuzov reaction. Thereafter reaction of dimethyl methylphosphonate with thionylchlorid yielded methylphosphonic dichloride. Butyllithium (2.5M in hexanes) was purchased from Merck and used as received. Diethylether was dried over sodium potassium alloy and distilled prior to use. Cyclohexanol was purchased from Merck, dried over sodium and distilled prior to use. Diethylamine was purchased from abcr, dried over CaH and distilled prior to use.
For substance preparation and characterisation, NMR spectra were measured with Varian 500VNMRS and Varian MR-400 spectrometers at 298\,K. 
Chemical shifts were referenced to residual protic impurities in the solvent ($^{1}$H) or the deuterio solvent itself ($^{13}$C) and reported relative to external SiMe$_{4}$ ($^{1}$H, $^{13}$C or 85$\%$ aqueous H$_{3}$PO$_{4}$ for $^{31}$P). 

\begin{figure*}
\centering
\includegraphics[width=7in]{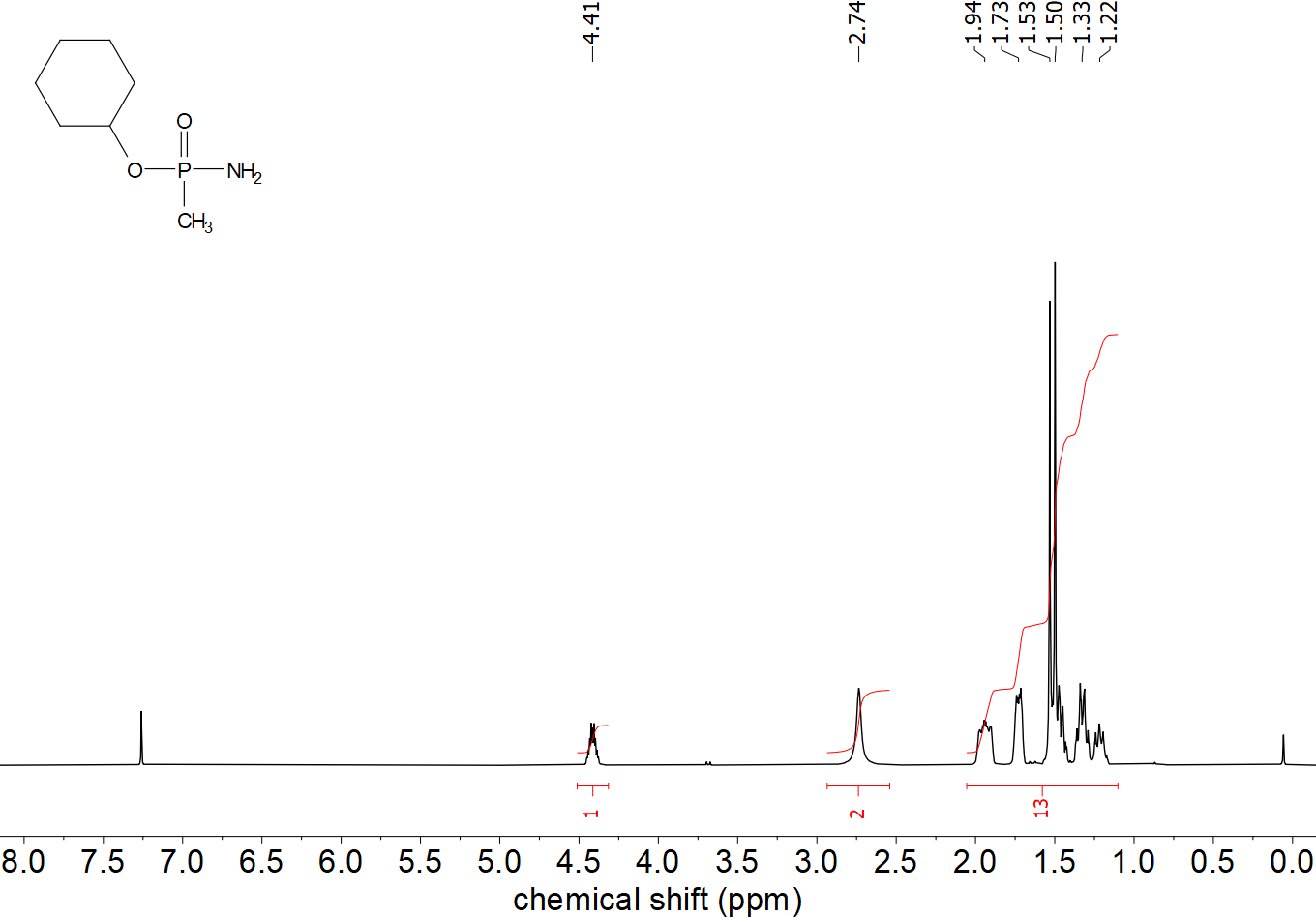}
\caption{$^{1}$H NMR spectrum (400\,MHz, CDCl$_{3}$) of Cyclohexyl \textit{P}-methylphosphonamidate. The methyl protons used to extract diastereomeric splitting are observed at 1.50 ppm and 1.53 ppm, split by coupling to phosphorus by a $J$-coupling of $16.8$ Hz. The residual \proton\ signal from the chloroform-d solvent at 7.4 ppm is used to reference the spectra.}

\end{figure*}

\begin{figure*}
\centering
\includegraphics[width=7in]{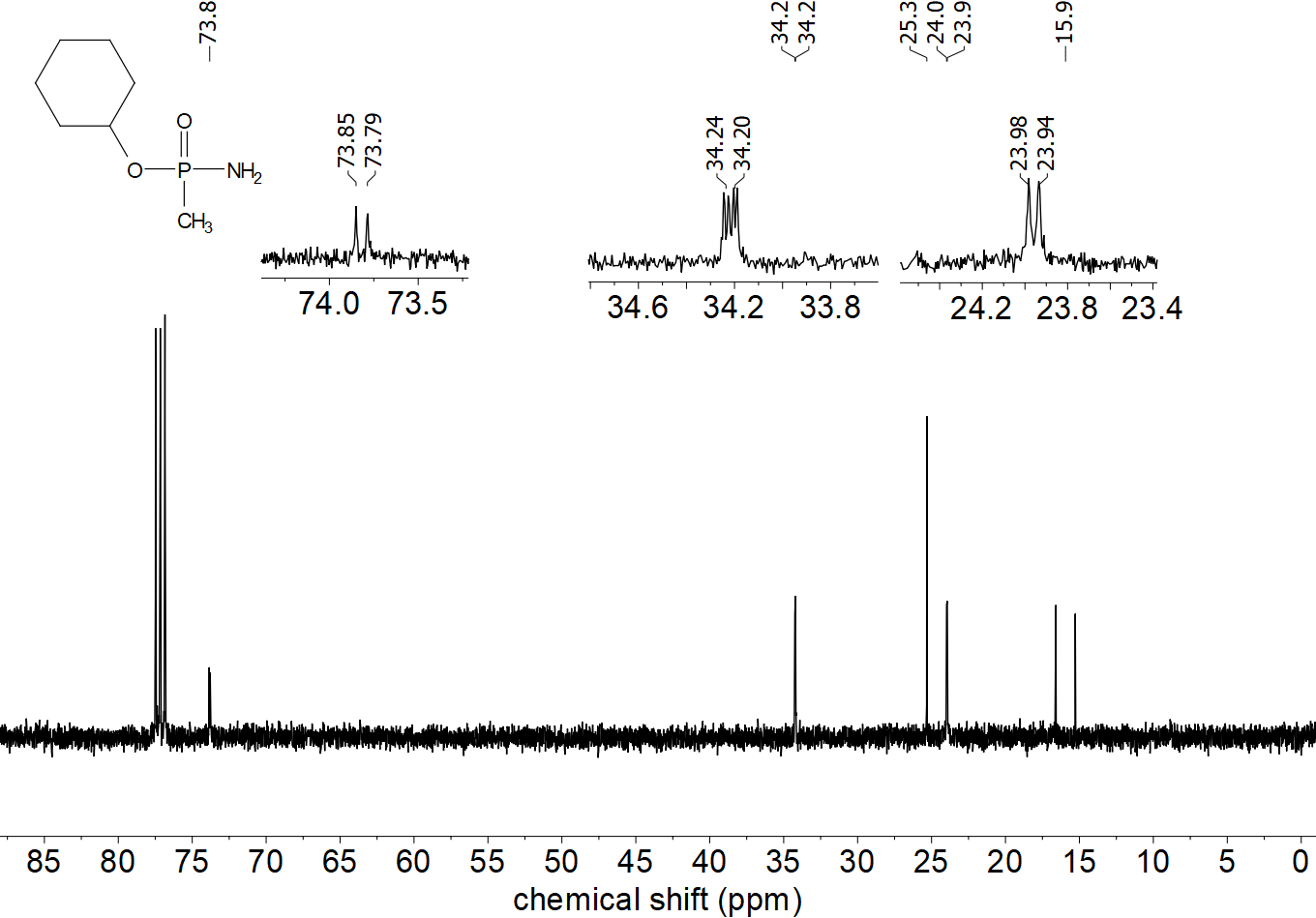}
\caption{$^{13}$C NMR spectrum (101\,MHz, CDCl$_3$) of Cyclohexyl \textit{P}-methylphosphonamidate with selected detail enlargement.}

\end{figure*}

\begin{figure*}
\centering
\includegraphics[width=7in]{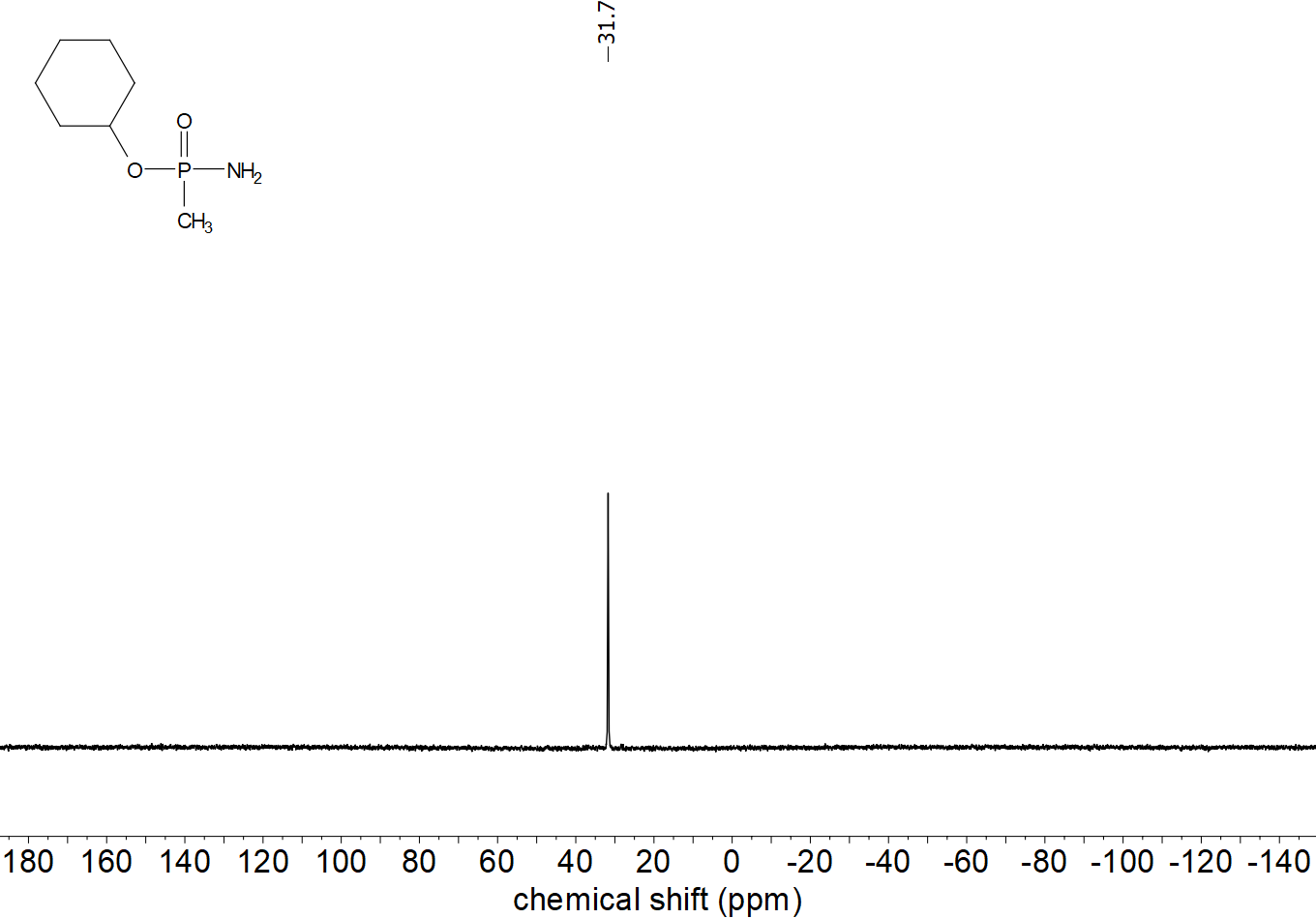}
\caption{$^{31}$P NMR spectrum (202\,MHz, CDCl$_{3}$) of Cyclohexyl \textit{P}-methylphosphonamidate.}

\end{figure*}

\begin{figure*}
\centering
\includegraphics[width=7in]{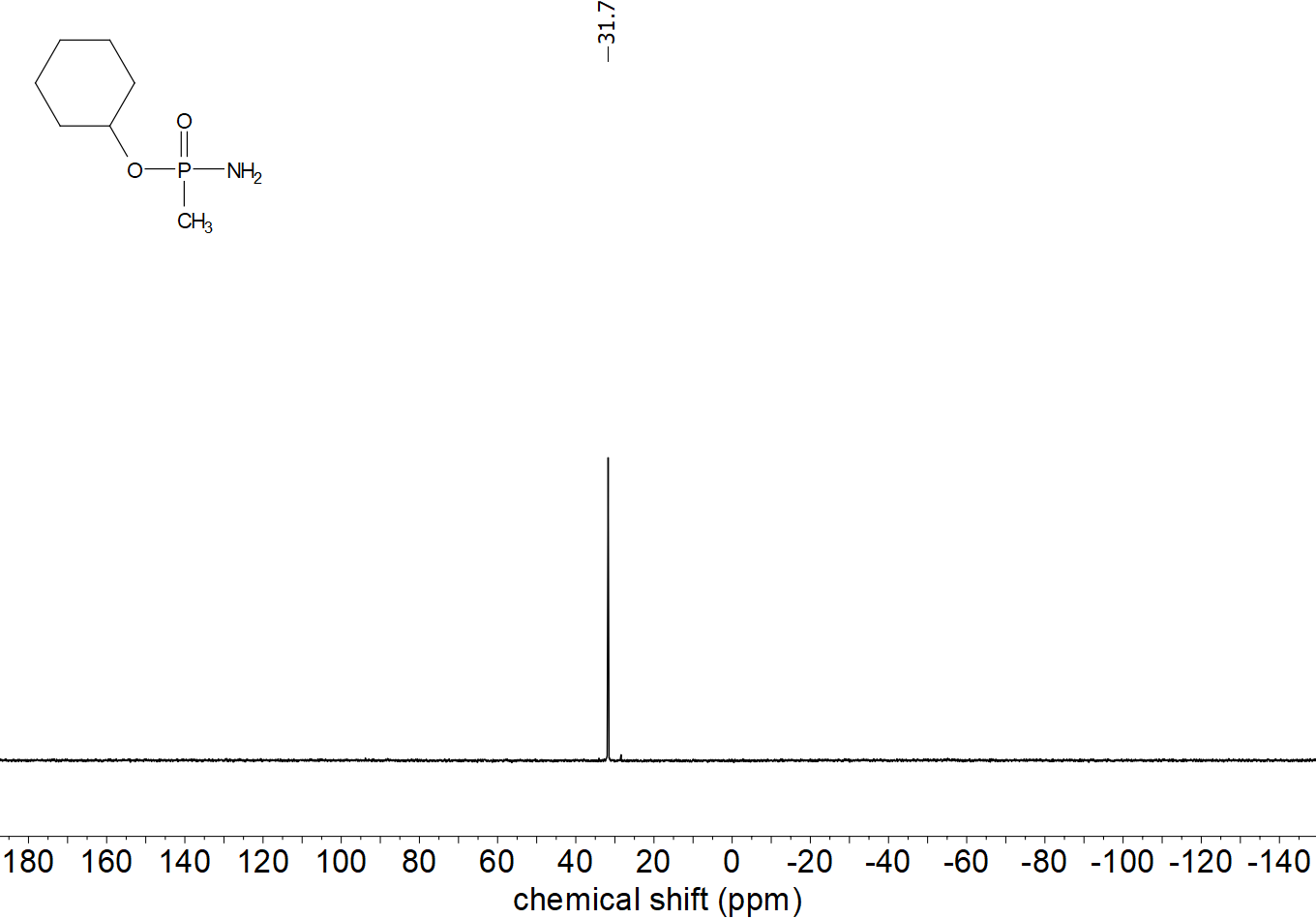}
\caption{$^{31}$P\{$^{1}$H\} NMR spectrum (202\,MHz, CDCl$_{3}$) of Cyclohexyl \textit{P}-methylphosphonamidate.}

\end{figure*}

\begin{figure*}
\centering
\includegraphics[width=7in]{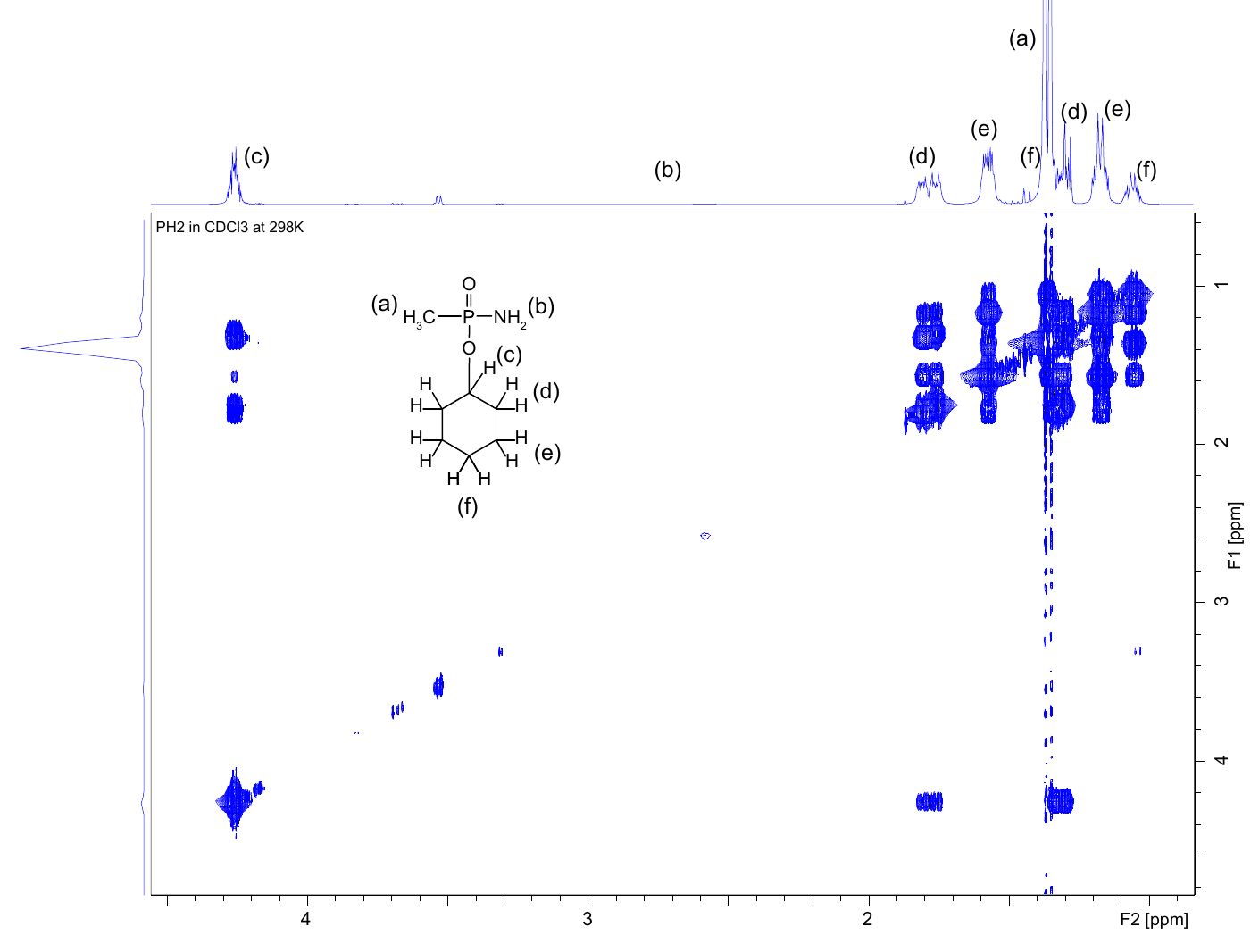}
\caption{$^{1}$H COSY NMR spectrum (850\,MHz, CDCl$_{3}$, 298\,K) of Cyclohexyl \textit{P}-methylphosphonamidate.}

\end{figure*}

\begin{figure*}
\centering
\includegraphics[width=7in]{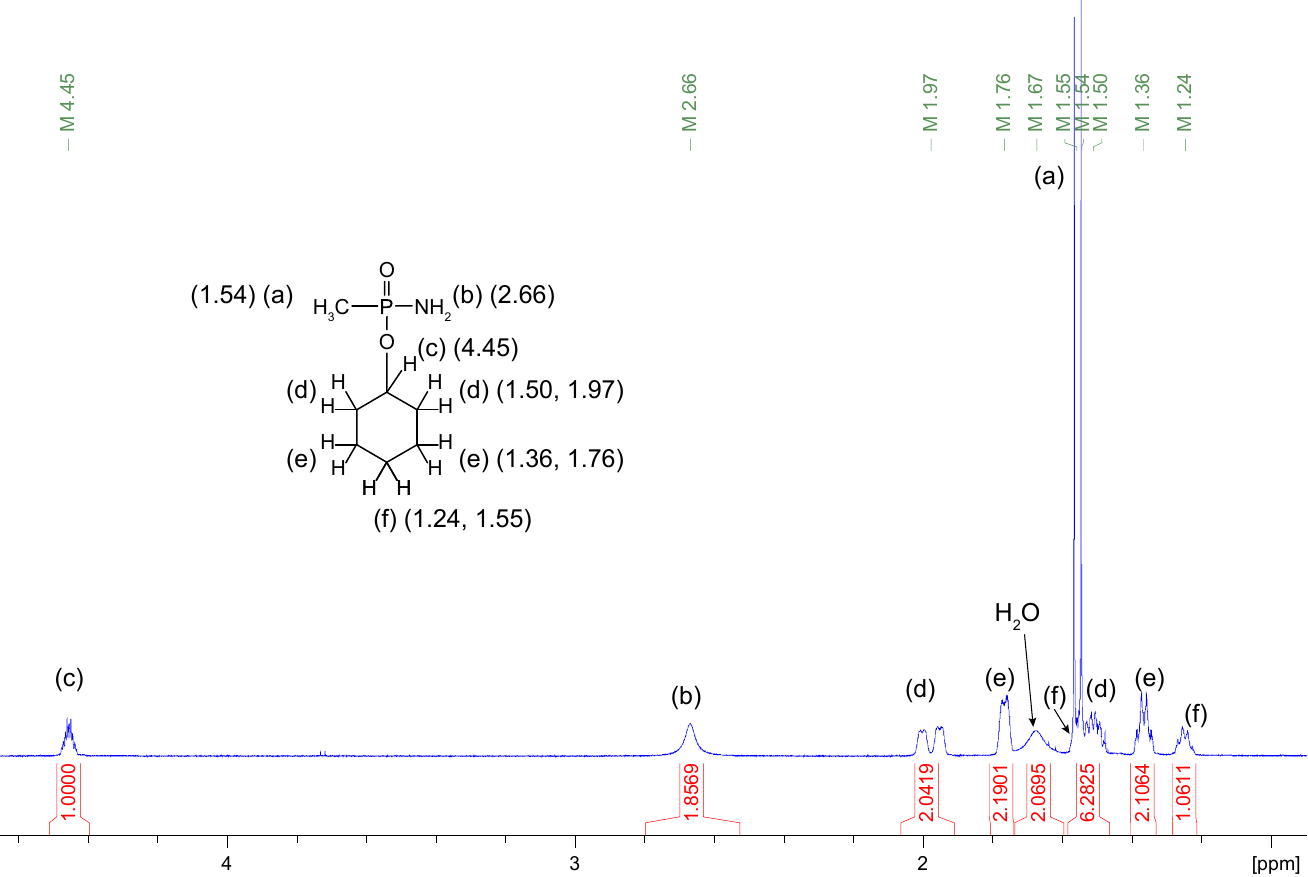}
\caption{$^{1}$H NMR spectrum (850\,MHz, CDCl$_{3}$, 298\,K) of Cyclohexyl \textit{P}-methylphosphonamidate. For ease of interpretation, only peak centers are indicated and splitting due to $J$-coupling is ignored.}

\end{figure*}

\FloatBarrier
\newpage
\section{Examination of diastereomeric splitting of \probe\ in the presence of chiral solvating agent}
\begin{figure*}
\begin{subfigure}{0.2\textwidth}
\centering
\includegraphics[width=7in]{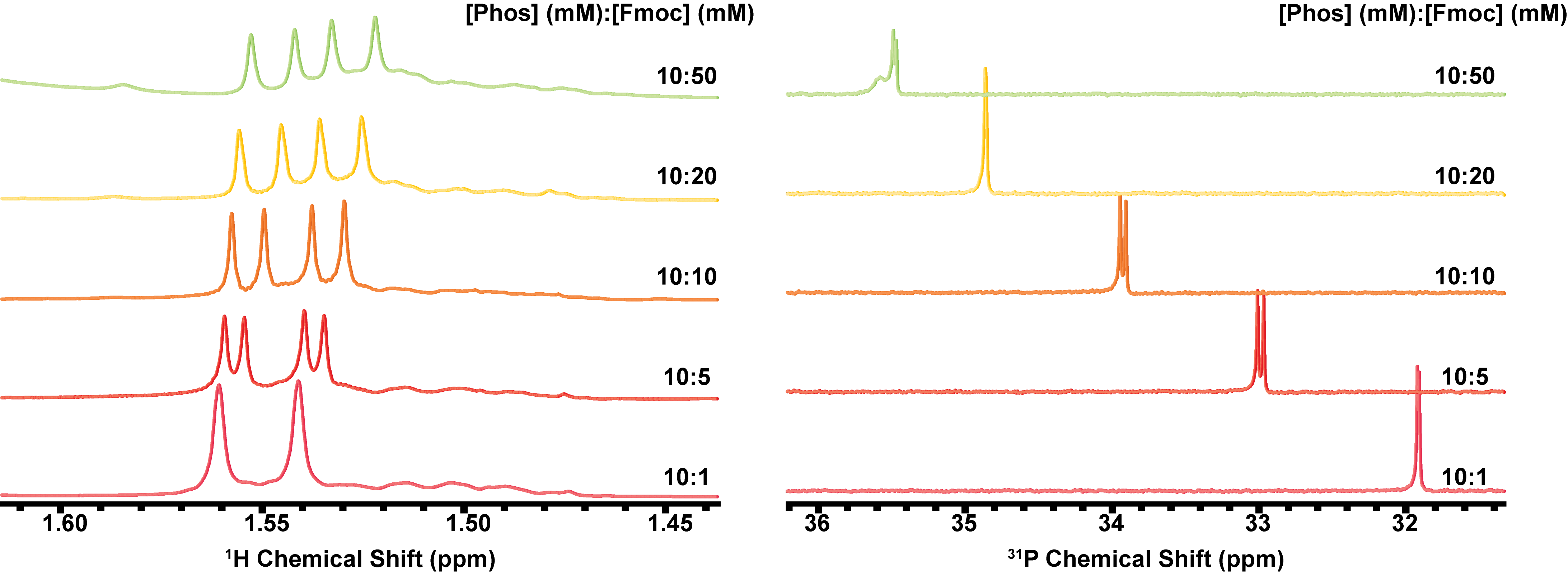}
\end{subfigure}

\begin{subfigure}{0.5\textwidth}
\includegraphics[width=7in]{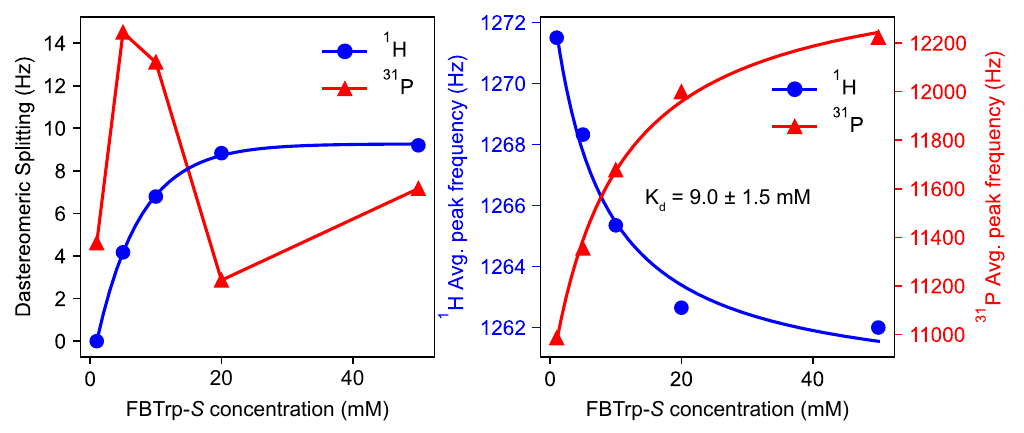}
\end{subfigure}
\caption{Increasing the ratio of \fmoc:\probe\ with 10\,mM \probe\ in chloroform-d. \proton\ spectra (left) shown are the sum of 32 scans and \phos\ spectra (right) are the sum of 64 scans collected with proton decoupling. While \proton\ diastereomeric splitting (\deltad) increases asymptotically to a maximum at \deltad $= 9.5$ Hz as the ratio of \fmocS:\probe\ is increased, \deltad\phos\ increases a point before collapsing at higher \fmocS:\probe. The ratio of 1:1 was chosen for the titration experiments in this study to maximize the total \deltad\ with the hopes of maximizing the total precision of the comagnetometry measurement. However since this ratio is outside of the linear regime in both \proton\ and \phos, changes in concentration of \probe\ or \fmoc\ are likely to cause shifts in \deltad\ that are not compensated by the comagnetometry approach. In future experiments, 
a ratio where changes in splitting of both nuclei are linear should be chosen. Binding affinity is determined from the shift in peak frequency of both \proton\ and \phos\ spectra using the model \[\delta_\mathrm{obs} = \delta_\mathrm{free} + (\delta_\mathrm{free} - \delta_\mathrm{bound})\frac{K_d}{[L]+K_d},\] where K$_d$ is the dissociation constant, $\delta_\mathrm{obs}$ is the observed chemical shift, $\delta_\mathrm{free}$ is the chemical shift of the free species, $\delta_\mathrm{bound}$ is the chemical shift of the bound species, and $[L]$ is the concentration of \fmocS. The reported $K_d$ is the weighted average of values given by fits of \proton\ ($7.0 \pm 2.9$ mM) and \phos\ ($10.0 \pm 1.8$ mM) average peak frequencies.}
\label{fgr:ratioSplittingStack} 
\end{figure*}

\begin{figure*}
\centering
\includegraphics[height=10cm]{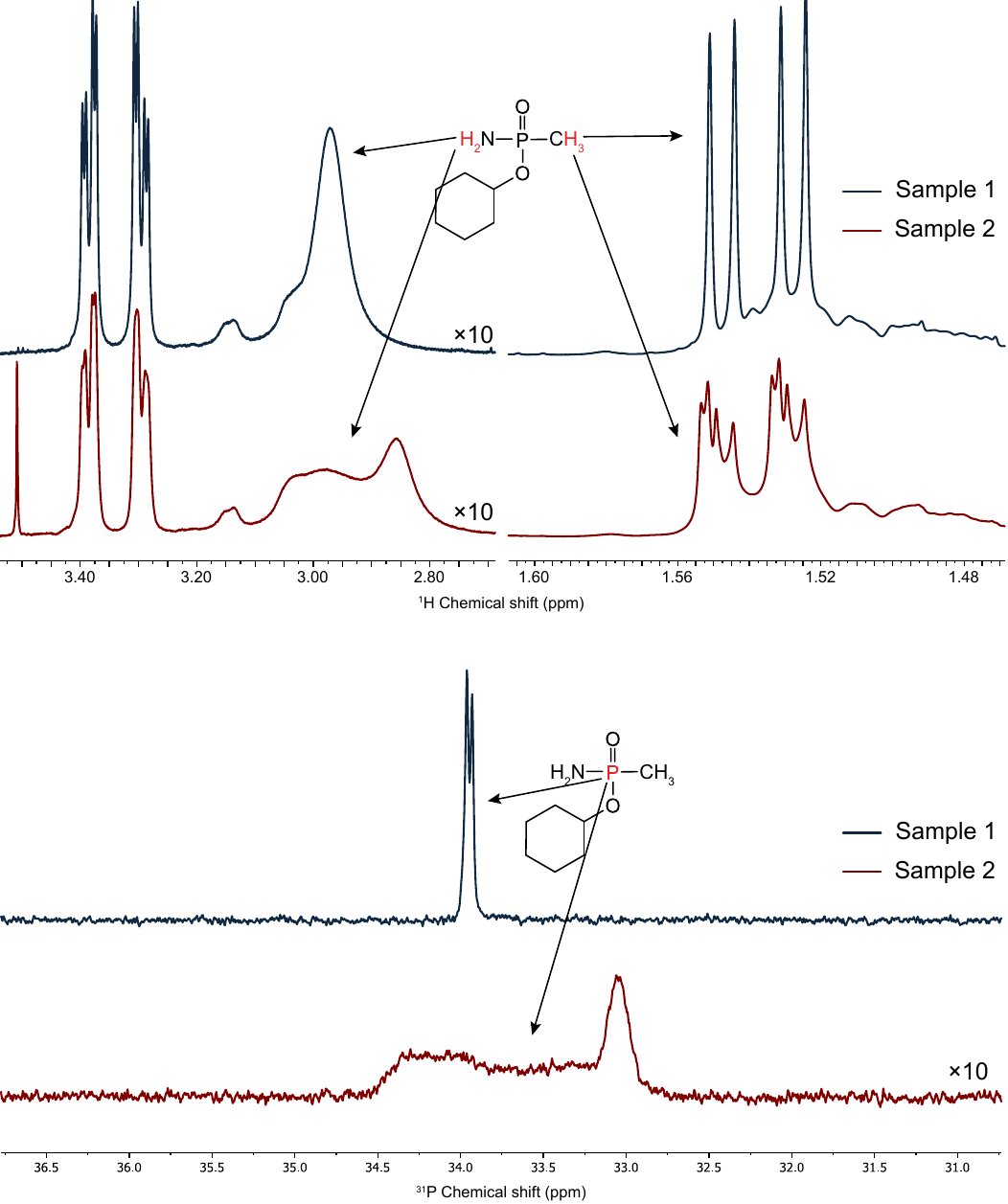}
\caption{Example of sample degradation observed in a sample with 10 mM \probe\ and \fmoc\ with an enantiomeric ratio of 6\% \fmocS\ in chloroform-d. The top \proton\ and \phos\ spectra are of a sample which did not display sample degradation over a 24 hour storage period at 4 \textdegree C. The bottom \proton\ and \phos\ spectra are of different sample with the same concentrations of \probe\ and \fmoc\ in the same chloroform-d solvent but from a different bottle, with clear signs of the formation of additional species due to unspecified chemical reactions.}
\label{fgr:SICorruptedSample}
\end{figure*}

\begin{figure*}
\centering
\includegraphics[height=2cm]{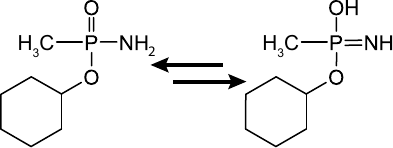}
\caption{Imine-amine tautomerization may occur in the probe molecule, creating the highly reactive imine. This may explain the sample degradation observed in some samples which was seemingly dependent on the storage and age of the solvent chloroform-d used to prepare the samples.}
\label{fgr:SItautomerization}
\end{figure*}

\begin{figure*}
\centering
\includegraphics[height=18cm]{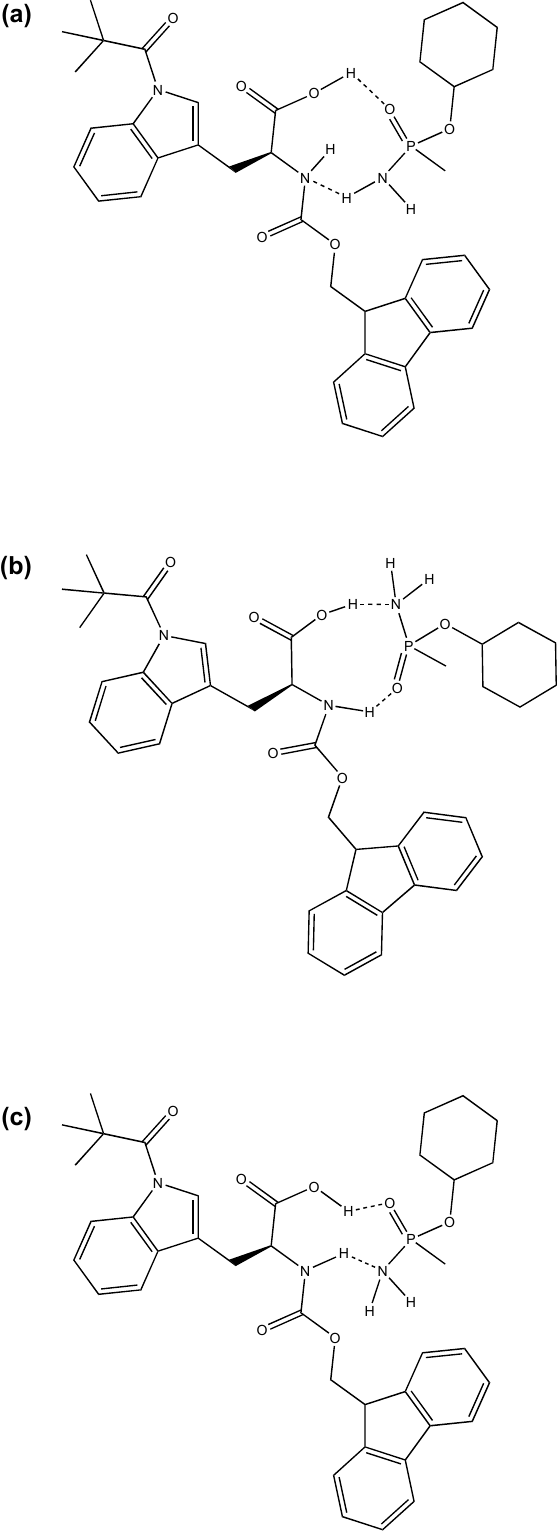}
\caption{Binding interactions between the chiral phosphonamidate \probe\ and substituted tryptophan \fmoc\ could be mediated by two groups on each molecule, namely the phosphoryl oxygen and amino moieties of \probe\ and the carboxylic acid and amino moieties of \fmoc. This work did not include characterization of this binding interaction, however the structures shown here depict likely binding interactions that may be occurring in our samples.}
\label{fgr:bindingInteractions}
\end{figure*}

\begin{figure}
\centering
  \includegraphics[width=8cm]{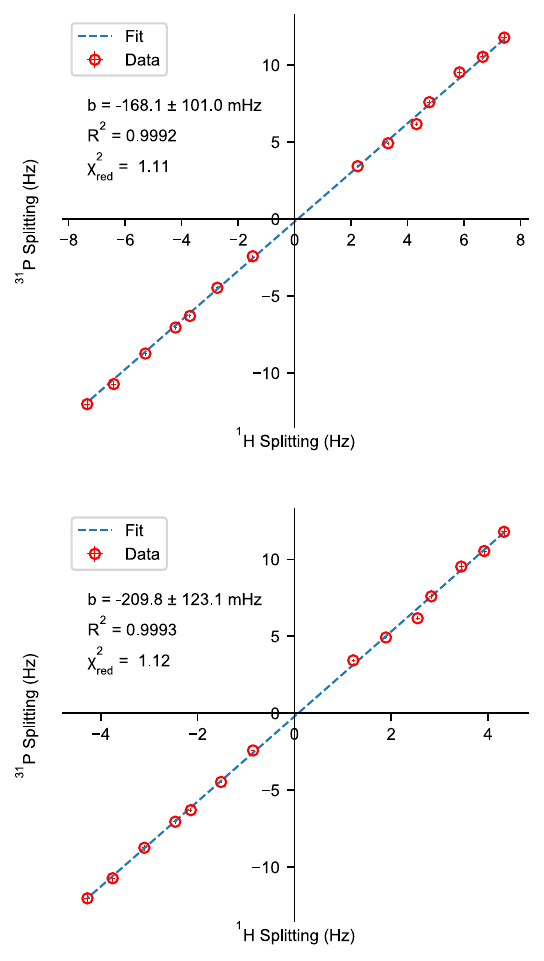}
  \caption{Comagnetometry plot showing diastereomeric splitting in spectra of \probe\ complexed with \fmoc\ as a function of \proton\ diastereomeric splitting two \proton\ multiplets. Both \phos\ and \proton\ spectra were collected using a 90\textdegree\ pulse, with \proton\ decoupling on \phos\ spectra. A reduced chi squared ($\chi_\mathrm{red}^{2}$) values of 1.12 and 1.11 indicate the error values used to compute the fit are a good estimation of the actual error, assuming a linear model.}
  \label{fgr:titration1}
\end{figure}

\begin{figure}
\centering
  \includegraphics[width=8cm]{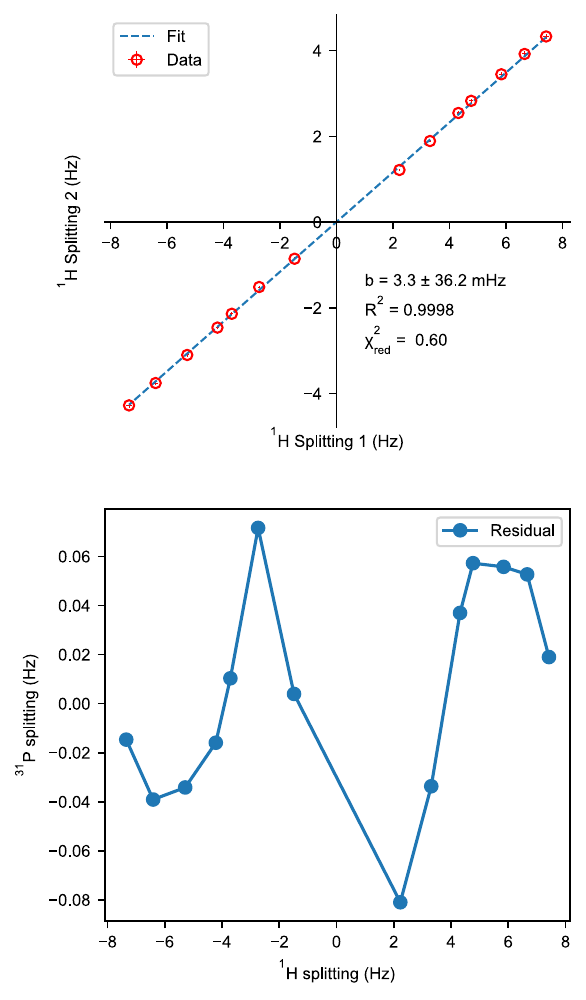}
  \caption{Comagnetometry plot showing correlation between diastereomeric splitting of two \proton\ multiplets originating from \probe. Both \proton\ \deltad\ values are extracted from spectra of the same samples and show a high degree of correlation. The plot below shows the difference between the fitted line and the measured values.}
  \label{fgr:protonvsproton}
\end{figure}

\begin{figure*}
\centering
  \includegraphics[width=3.3in]{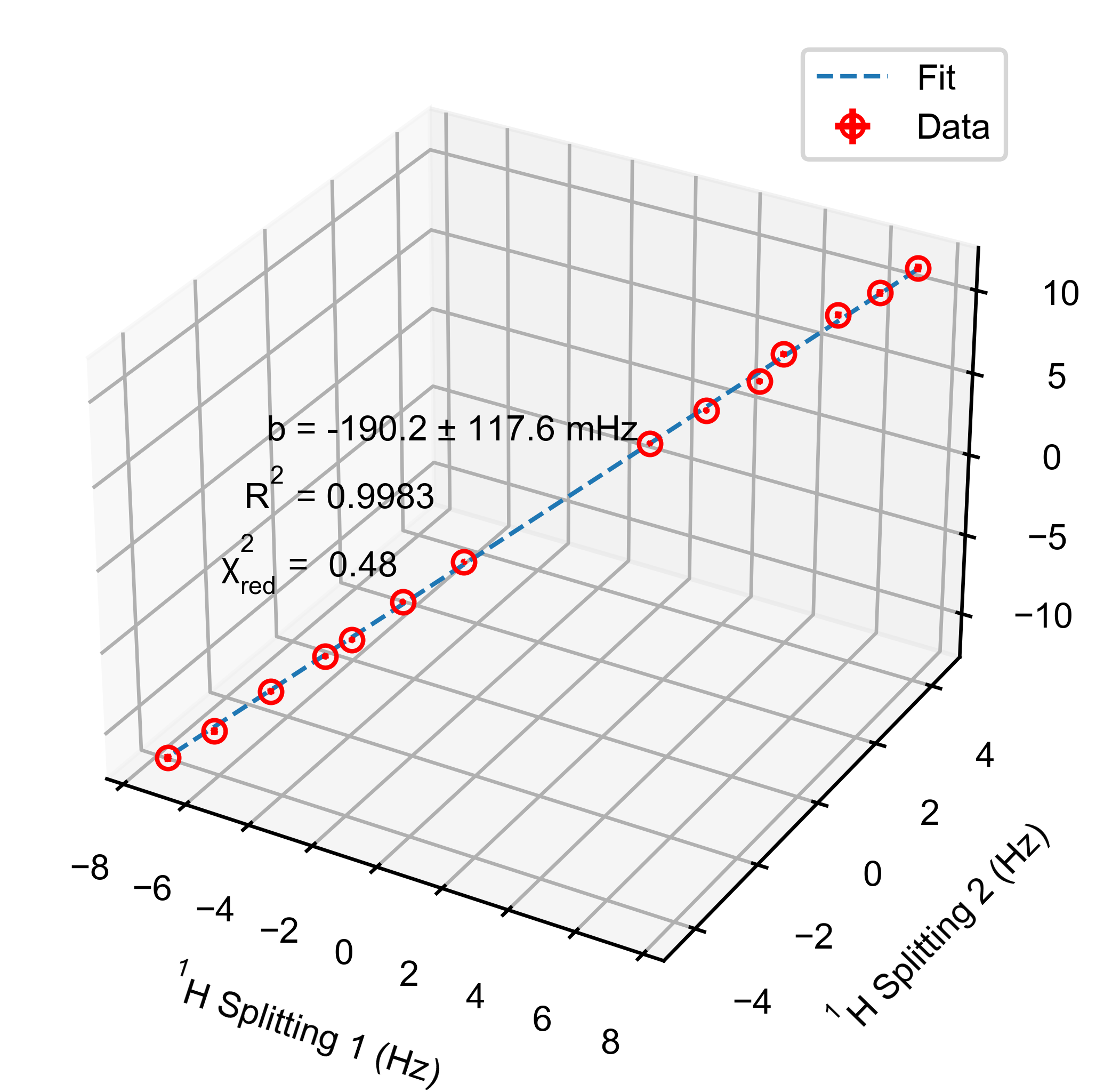}
  \caption{Plot of \phos\ splitting as a function of two \proton\ multiplets originating from \probe\ which display diastereomeric splitting. Fitting was accomplished by minimizing $\chi^2=\sum_i(\frac{y_{obs}-y_{x_i}}{T.E.})^2$ with $T.E.=\sqrt{(a\sigma_x)^2+(b\sigma_y)^2+\sigma_c^2+2a\rho_{xz}\sigma_x\sigma_y+2b\rho_{yz}\sigma_y\sigma_z+2ab\rho_{xy}\sigma_x\sigma_y}$ for the function $z=ax+by+c$}
  \label{fgr:3Dcomag}
\end{figure*}

\begin{figure*}
\centering
  \includegraphics[width=3.3in]{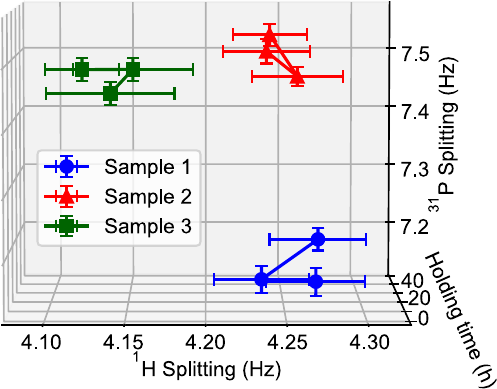}
  \caption{Reproducibility and time-stability of diastereomeric splitting samples prepared in an identical manner. Three samples were prepared with identical ratios of \fmocS:\fmocR\ (80:20, 10\,mM total concentration) and 10\,mM \probe\ and measured using the same conditions as the samples used to construct the comagnetometer plot (figure\,\ref{fgr:comagPlot}). From this it is clear that some variability of diastereomeric splitting is caused by sample preparation, possibly due to small differences in concentration. Fluctuations in \deltad\ over the 48\,h observation period are within the margin of error.}
  \label{fgr:timeDependence}
\end{figure*}

\begin{figure*}
\centering
\includegraphics[height=10cm]{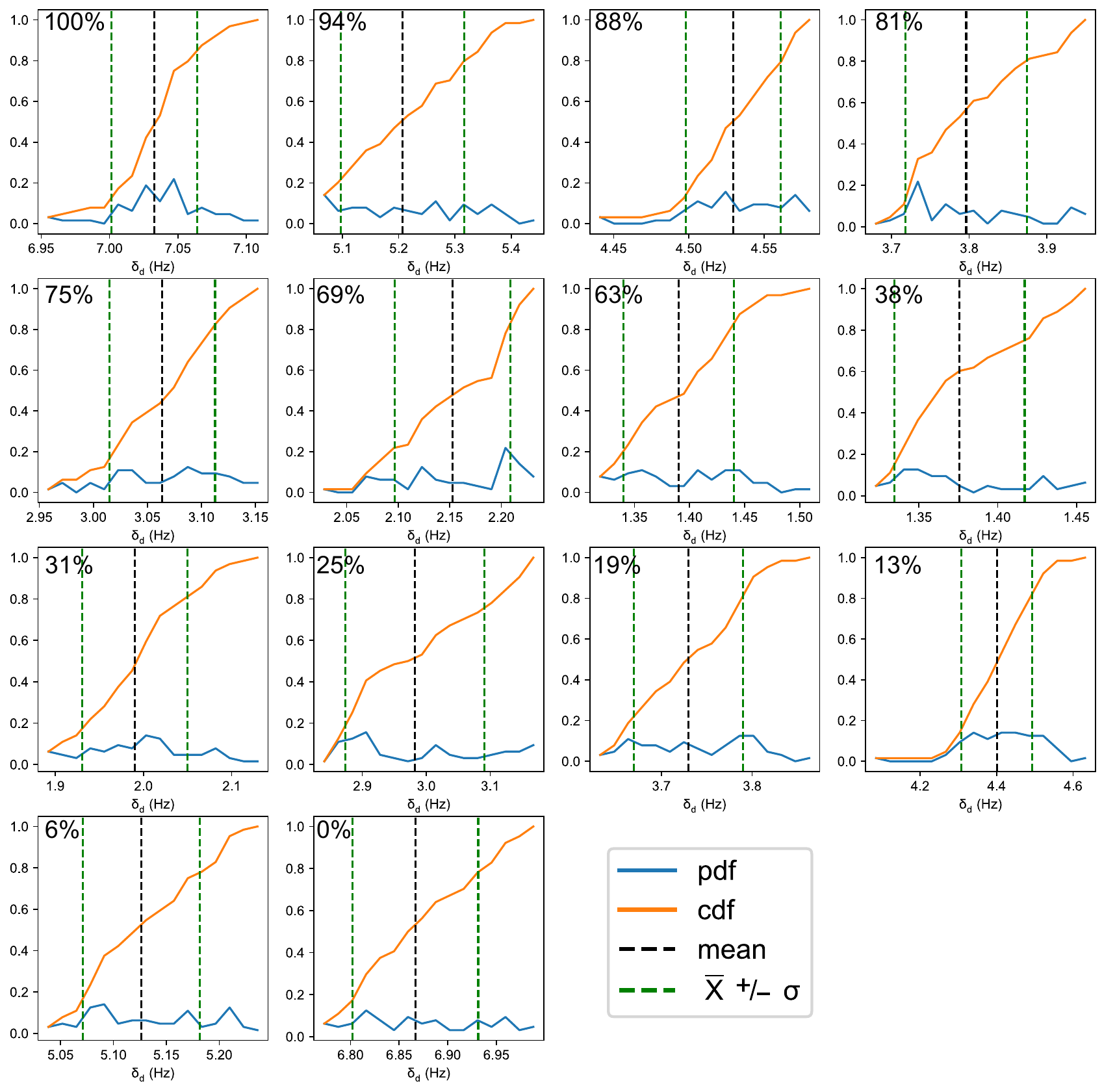}
\caption{Depiction of the probability distribution function (pdf, \textcolor{blue}{blue}) and cumulative distribution function (cdf, \textcolor{orange}{orange}) of diastereomeric splitting estimates from 64 \proton\ spectra of each sample used to construct the second comagnetometer plot with bilinear rotation decoupling (BIRD). Deviations from a normal (Gaussian) distribution means that the data cannot be well described by basic statistical parameters like standard deviation, and could indicate that systematic errors are present in the measurement.}
\label{fgr:SICDFandPDF}
\end{figure*}

\begin{figure*}
\centering
\includegraphics[height=10cm]{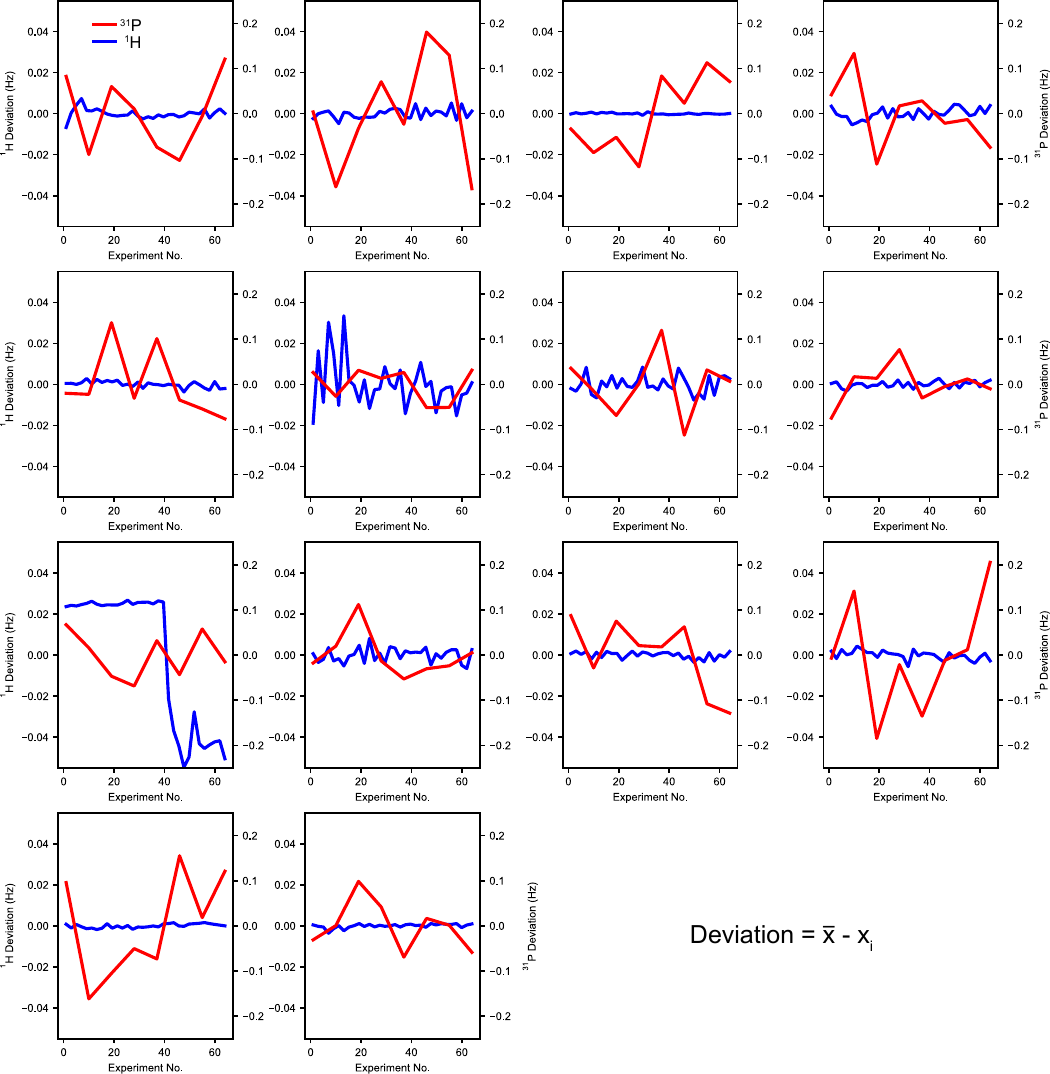}
\caption{Deviation of \deltad\ from the mean extracted by fitting \proton\ and \phos\ (without \proton\ decoupling) NMR spectra. \proton\ and \phos\ spectra were summed in groups of 2 and 8 respectively due to differences in signal-to-noise ratios which were lower for \phos\ spectra. Clearly, there is larger variance in \deltad\ extracted from \phos\ spectra and it is likely that obtaining more spectra of less-sensitive nuclei (compared to \proton) would help improve the confidence levels of fitting estimates. The systematic drift seen in \proton\ \deltad\ in the 3rd row, 1st column figure shows what may be a drift due to change in lineshape due to magnetic field homogeneity.}
\label{fgr:SIDeviation}
\end{figure*}

\FloatBarrier
\newpage
\section{Fitting of proton and phosphorus NMR spectra}

\begin{figure*}
\centering
\includegraphics[width=3.3in]{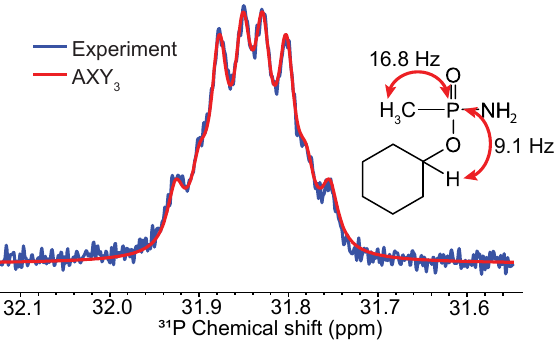}
\caption{Spectrum of chiral phosphorus probe with overlay of simulated line shape for an AXY\textsubscript{3} system. The trace in red is the result of fitting using ANATOLIA \cite{Cheshkov2018} which was used to estimate $J$-couplings for the spin system and spectral line-widths to generate a fitting function to extract frequency estimates from non-\proton\ decoupled \phos\ spectra.}
\label{fgr:AnatoliaFit}
\end{figure*}

\proton\ spectra were fit in python using the $curve\_fit$ package from $scipy.optimize$. A sum of complex Lorentzians with the general form 

\[\sum_{i}^{4} a_i \left( \frac{\Gamma^2}{\Gamma^2+(\nu-\nu_{0i})^2}\cos(\phi_i) + \frac{\nu-\nu_{0i}}{\Gamma^2+(\nu-\nu_{0i})^2} \sin(\phi_i) \right) \]

was used as the fitting function. Note that the amplitude ($a$), center frequency ($\nu_0$), and phase ($\phi$) are independent for each Lorentzian, while the width ($\Gamma$) is set to be the same for all peaks. This model generated the best fits judging by the residuals and by the variance estimates produced by the fitting algorithm. A sum of two Lorentzian doublets was also used but produced slightly larger error despite having fewer parameters compared to the four Lorentzian model.

\begin{figure*}
\centering
\includegraphics[height=15cm]{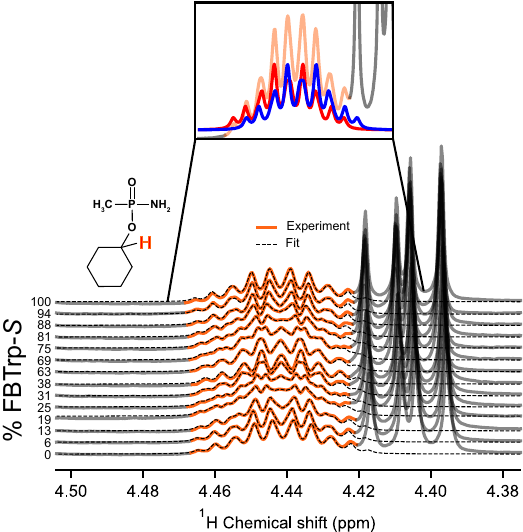}
\caption{\proton\ spectra of the second multiplet from \probe\ which displays diastereomeric splitting. This multiplet originates from a single cyclohexyl proton nearest the \phos\ center, highlighted in \textcolor{orange}{orange}. \deltad\ was extracted by fitting two analytical multiplet functions shown in the inset based on an assumed AM$_2$N$_2$X spin system.}
\label{fgr:SIcyclohexylmultiplet}
\end{figure*}

\begin{figure*}
\centering
\includegraphics[width=7in]{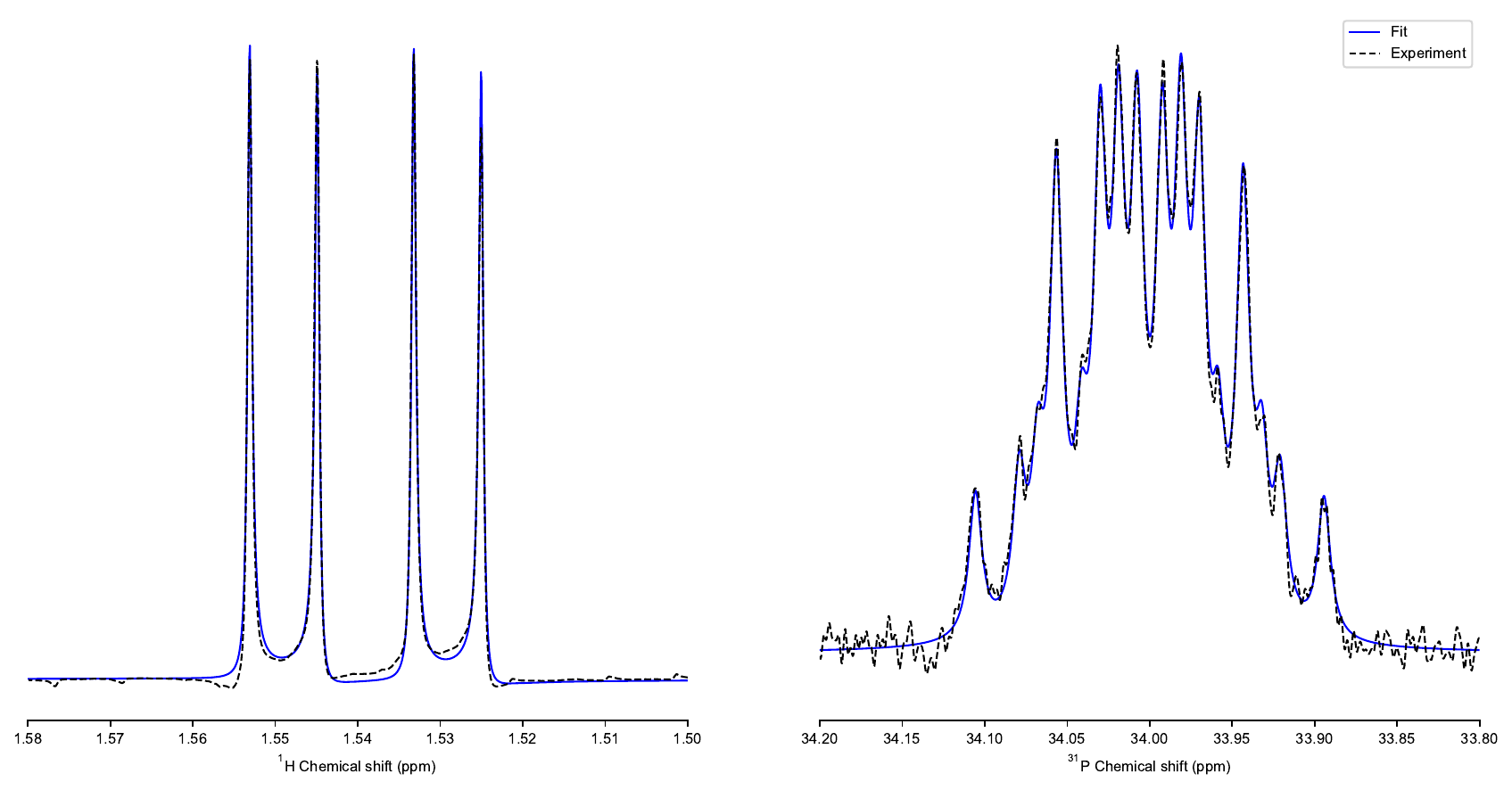}
\caption{Example plots of experimental data fitted with Lorentzian functions to extract diastereomeric splitting.}
\label{fgr:SIfitExample}
\end{figure*}

\FloatBarrier
\newpage
\section{BIRD pulse sequence}
\begin{figure*}
\centering
\includegraphics[height=5cm]{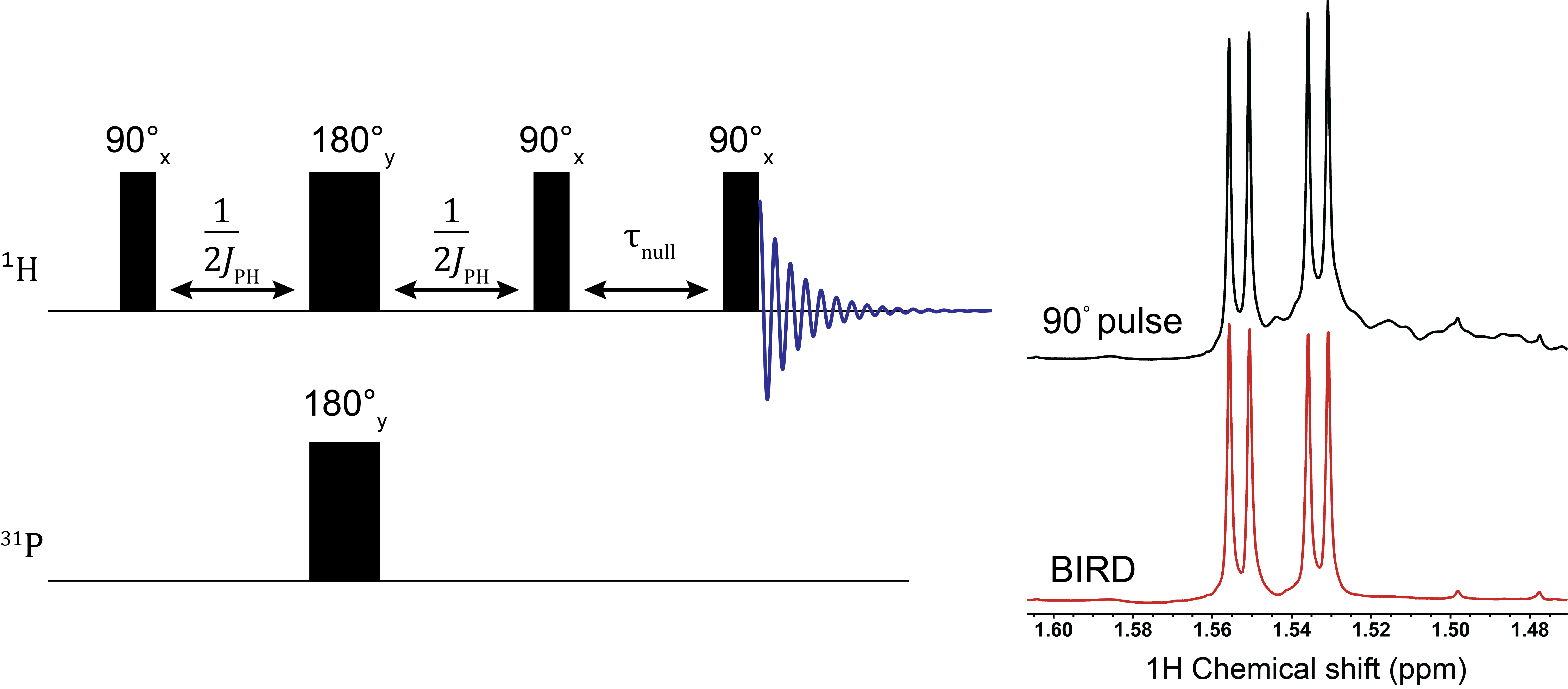}
\caption{Graphical depiction of the bilinear rotational decoupling (BIRD) pulse sequence used in proton spectral acquisition in the second titration. The object of the pulse sequence in this context is to orient all proton spins not coupled to \phos\ against the applied magnetic field while leaving the spins coupled to \phos\ along the applied field. A short delay before the final readout pulse is set to allow all spins aligned against the field to relax to zero, thus removing broad peaks overlapping with the peaks of interest and allowing more precise modeling. Here, ${}^{2}J_\mathrm{PH}$ is the $J$-coupling between the methyl protons and phosphorus in the chiral probe molecule and $\tau_\mathrm{null}$ is the waiting period to allow the $z$-component of spins not coupled to \phos\ to relax until close to zero. Pulse sequences used ${}^{2}J_\mathrm{PH} = 16.7$ Hz and $\tau_\mathrm{null} = 100 $ ms.}
\label{fgr:SIBIRD}
\end{figure*}

\FloatBarrier

\newpage
\section{N,N-diethyl phosphorus probe molecule}

\begin{figure*}
\centering
\includegraphics[width=2in]{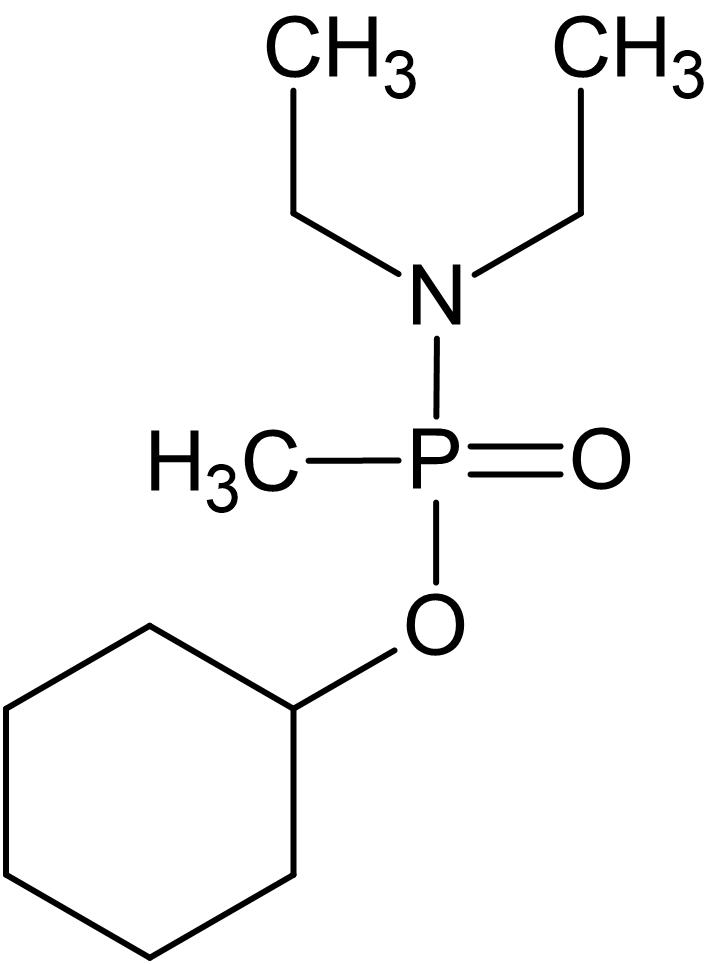}
\caption{Chemical structure of second chiral phosphorus compound examined in this study, cyclohexyl N,N-diethyl-\textit{P}-methylphosphonamidate (\DEprobe). The ethyl groups substituted on the amino group are expected to reduce reactivity of the compound with regards to the protonated amino form. However, the conditions tested failed to produce diastereomeric splitting in \phos\ spectra. Diastereomeric splitting was observed in \proton\ spectra under several concentrations and ratios of \DEprobe\ and \fmoc .}
\label{fgr:SIethylatedPhosStructure}
\end{figure*}

\begin{figure*}
\centering
\includegraphics[width=7in]{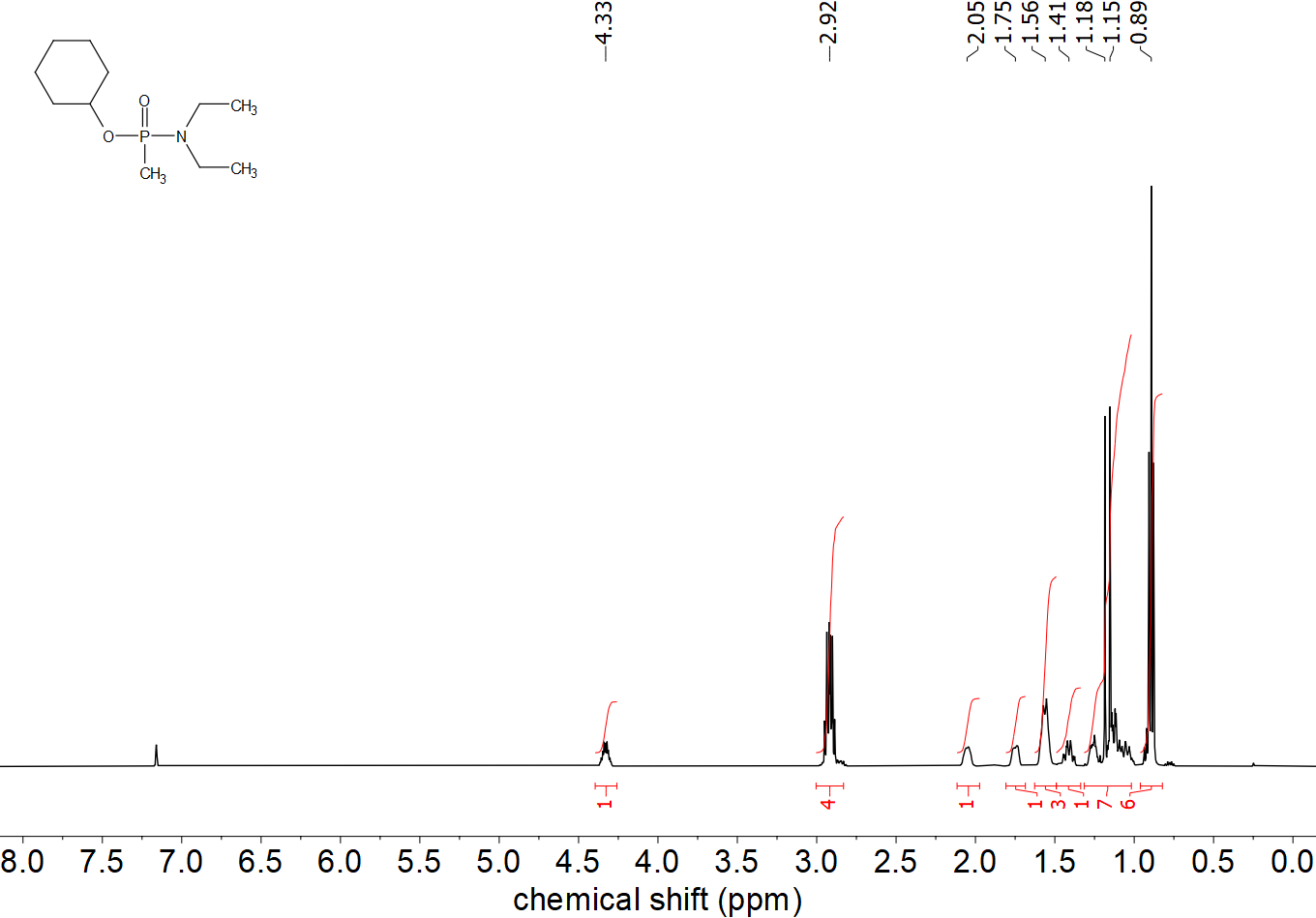}
\caption{$^{1}$H NMR spectrum (500\,MHz, C$_{6}$D$_{6}$) of Cyclohexyl \textit{N},\textit{N}-diethyl-\textit{P}-methylphosphonamidate.}

\end{figure*}

\begin{figure*}
\centering
\includegraphics[width=7in]{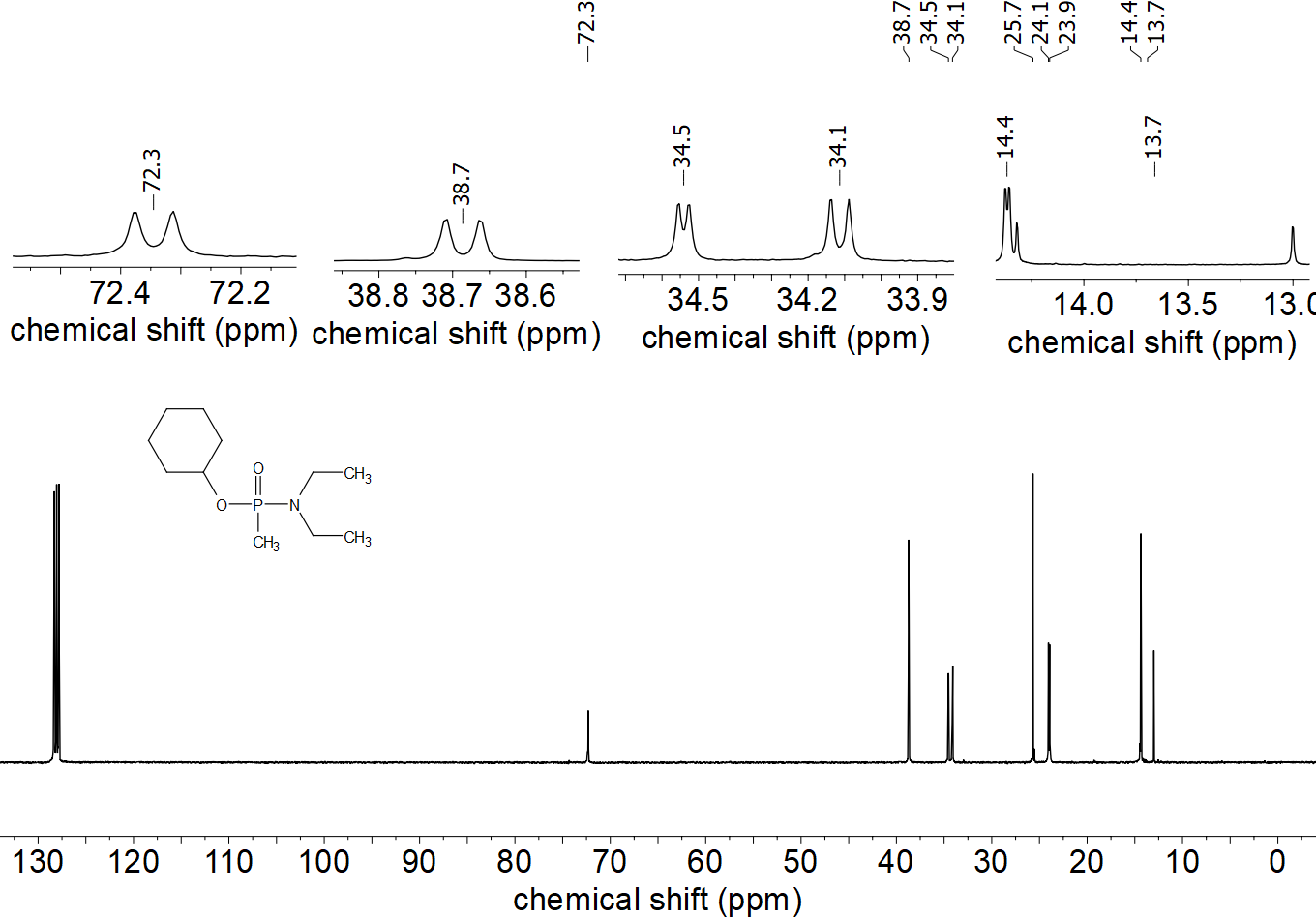}
\caption{$^{13}$C NMR spectrum (101\,MHz, C$_{6}$D$_{6}$) of Cyclohexyl \textit{N},\textit{N}-diethyl-\textit{P}-methylphosphonamidate with selected detail enlargement.}

\end{figure*}

\begin{figure*}
\centering
\includegraphics[width=7in]{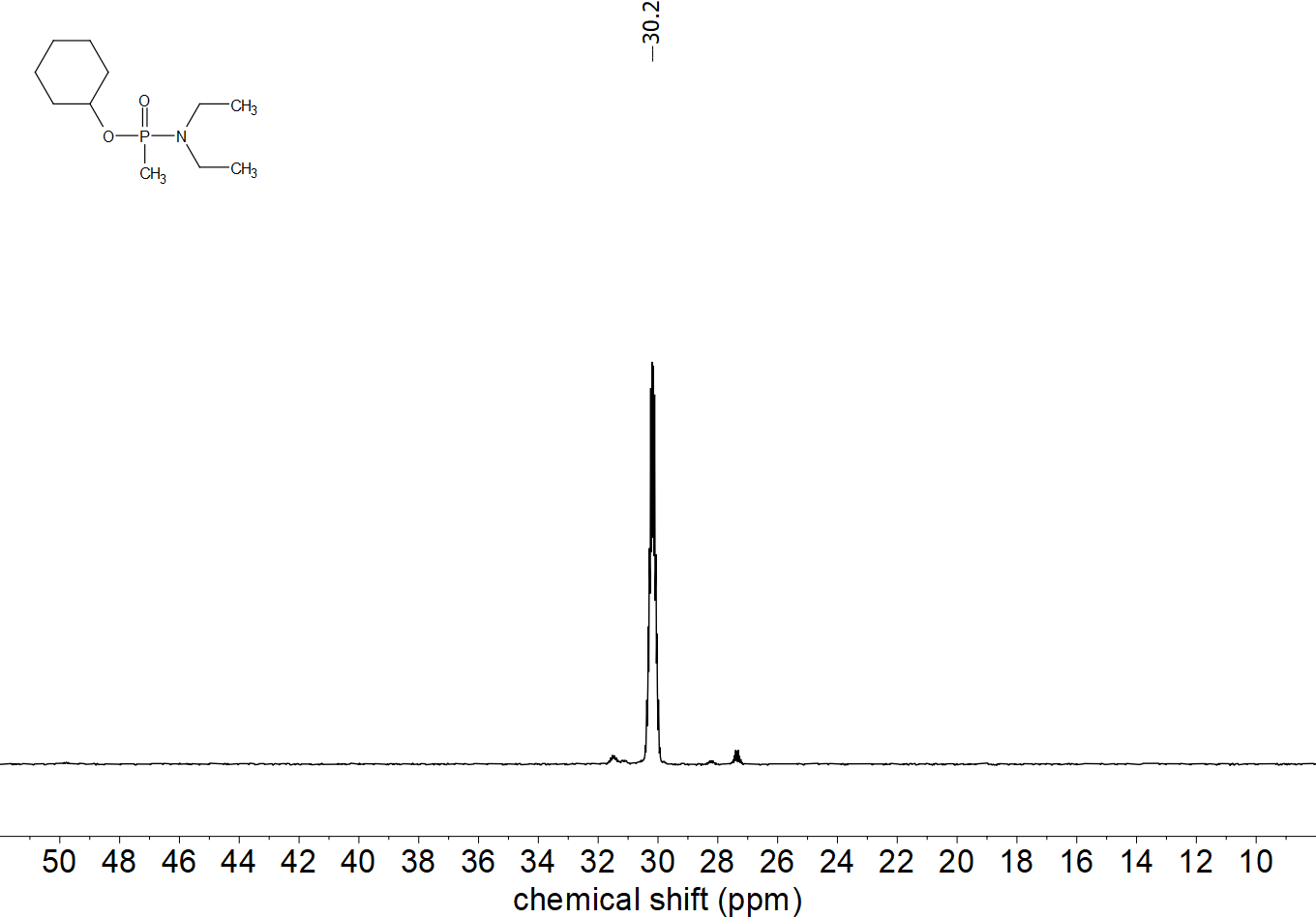}
\caption{$^{31}$P NMR spectrum (202\,MHz, C$_{6}$D$_{6}$) of Cyclohexyl \textit{N},\textit{N}-diethyl-\textit{P}-methylphosphonamidate.}

\end{figure*}

\begin{figure*}
\centering
\includegraphics[width=7in]{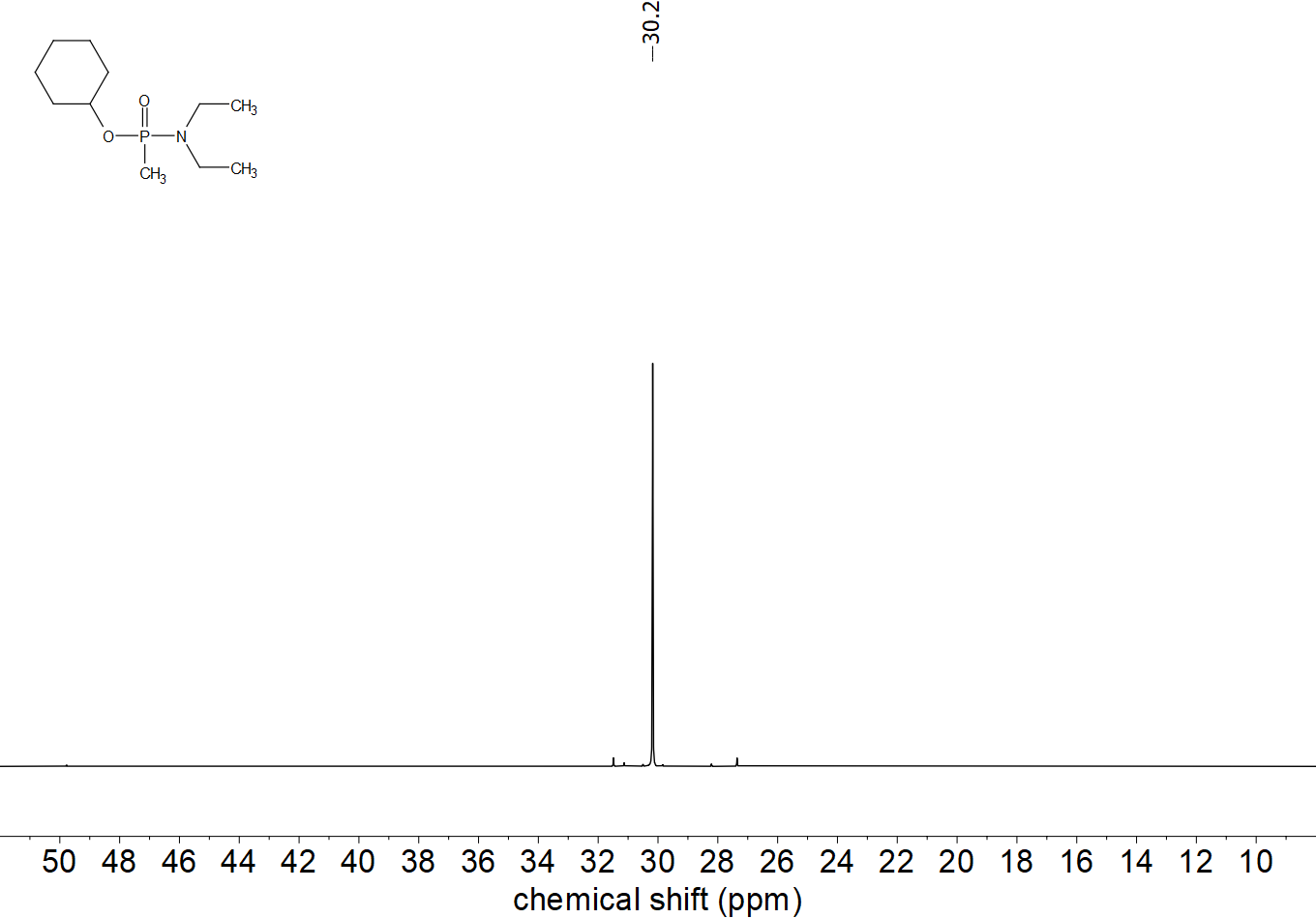}
\caption{$^{31}$P\{$^{1}$H\} NMR spectrum (202\,MHz, C$_{6}$D$_{6}$) of Cyclohexyl \textit{N},\textit{N}-diethyl-\textit{P}-methylphosphonamidate.}

\end{figure*}

\begin{figure*}
\centering
\includegraphics[width=7in]{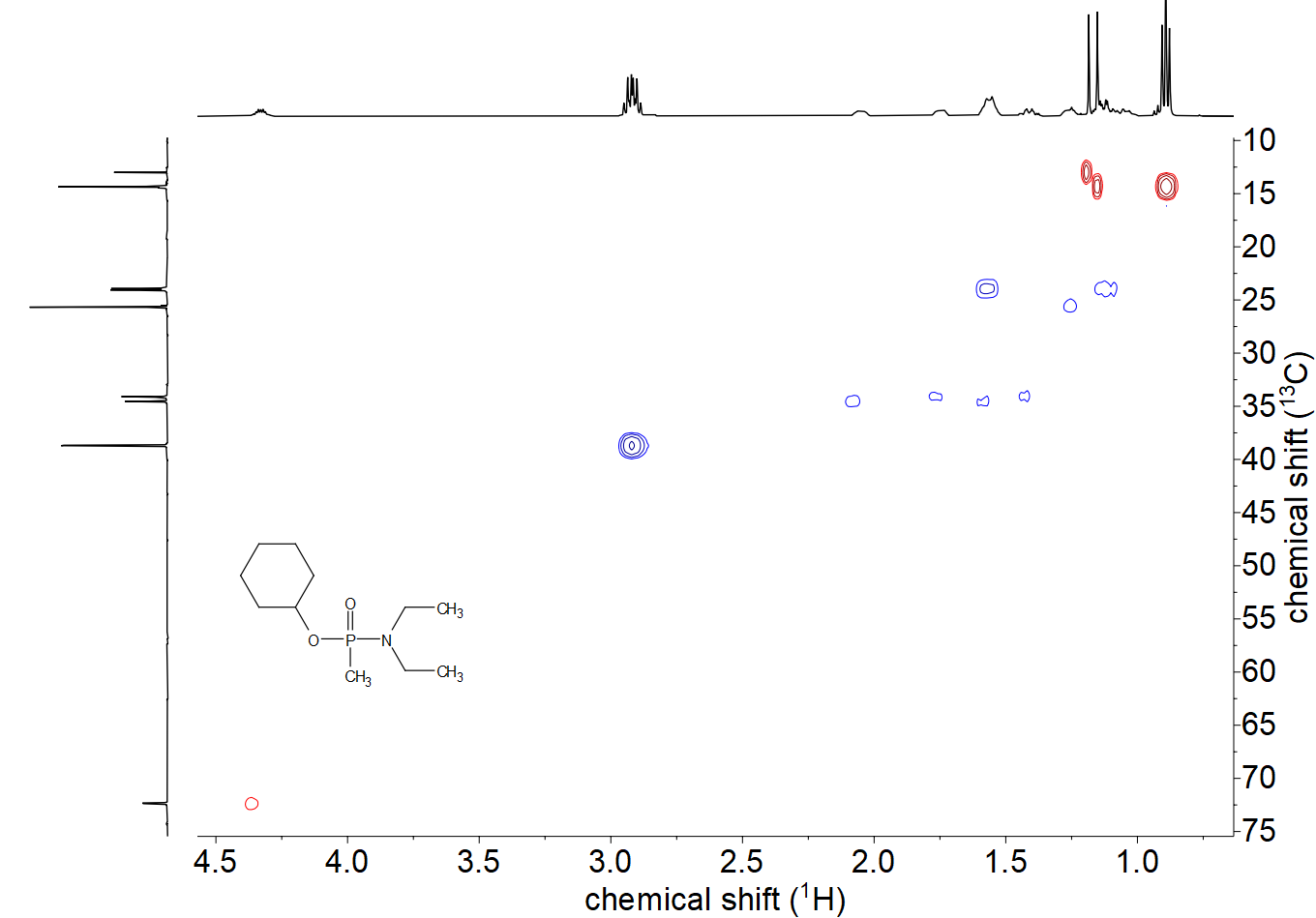}

\caption{2D-HSQC NMR spectrum (101\,MHz, C$_{6}$D$_{6}$) of Cyclohexyl \textit{N},\textit{N}-diethyl-\textit{P}-methylphosphonamidate.}
\end{figure*}

\begin{figure*}
\centering
\includegraphics[width=3.3in]{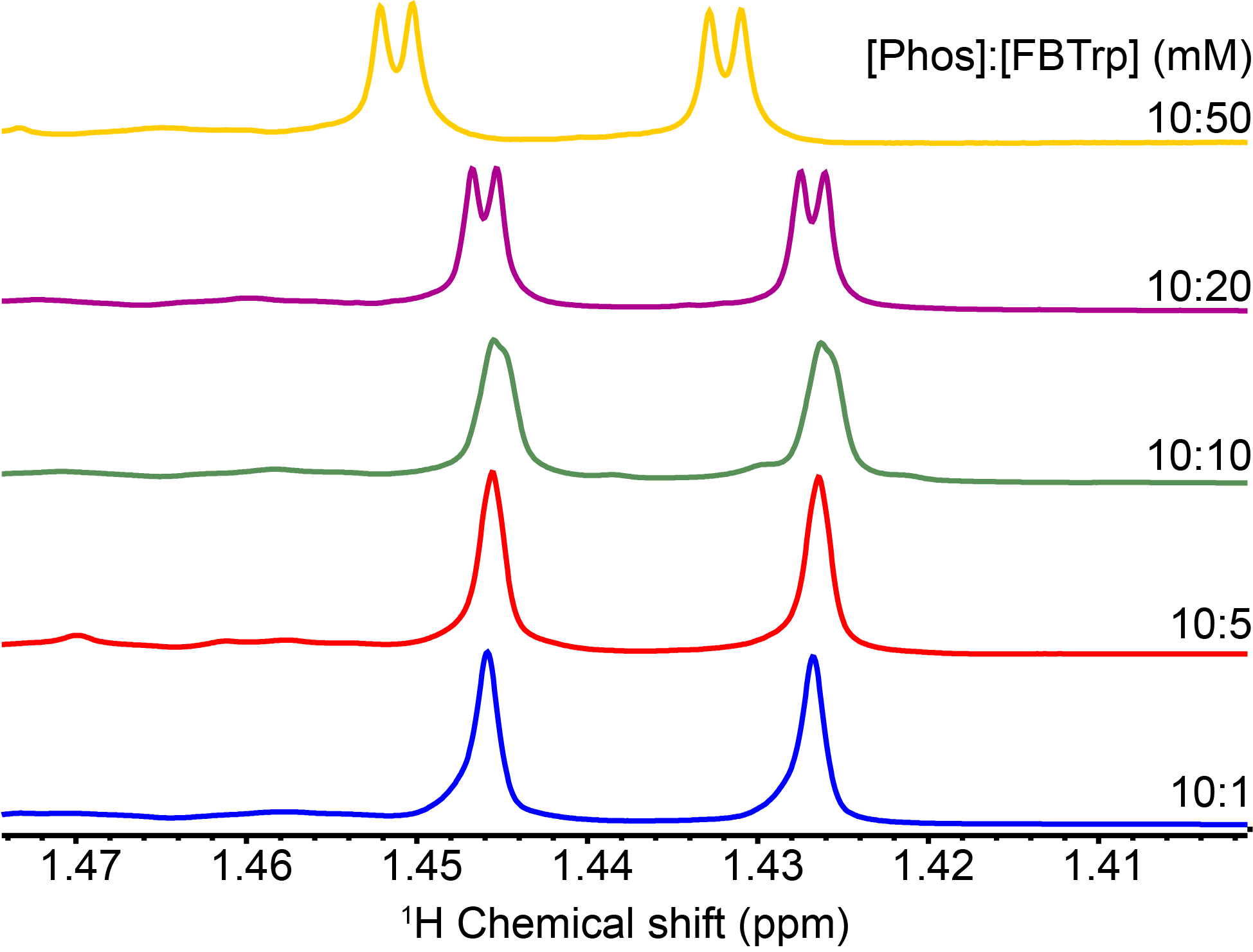}
\caption{Proton spectra at 20\,T and 298\,K of samples containing N,N-diethyl phosphorus chiral probe (\DEprobe) and \fmocS\ at several ratios. The concentration of \DEprobe\ was maintained constant at 10 mM while concentrations of \fmocS\ varied from 1 mM to 50 mM. Samples were prepared by mixing stock solutions of \DEprobe\ and \fmocS\ in chloroform directly in 5 mm NMR tubes and were allowed to equilibrate at room temperature for at least 1\,hr.}
\label{fgr:SIEthylPhosRatioProton}
\end{figure*}

\begin{figure*}
\centering
\includegraphics[width=3.3in]{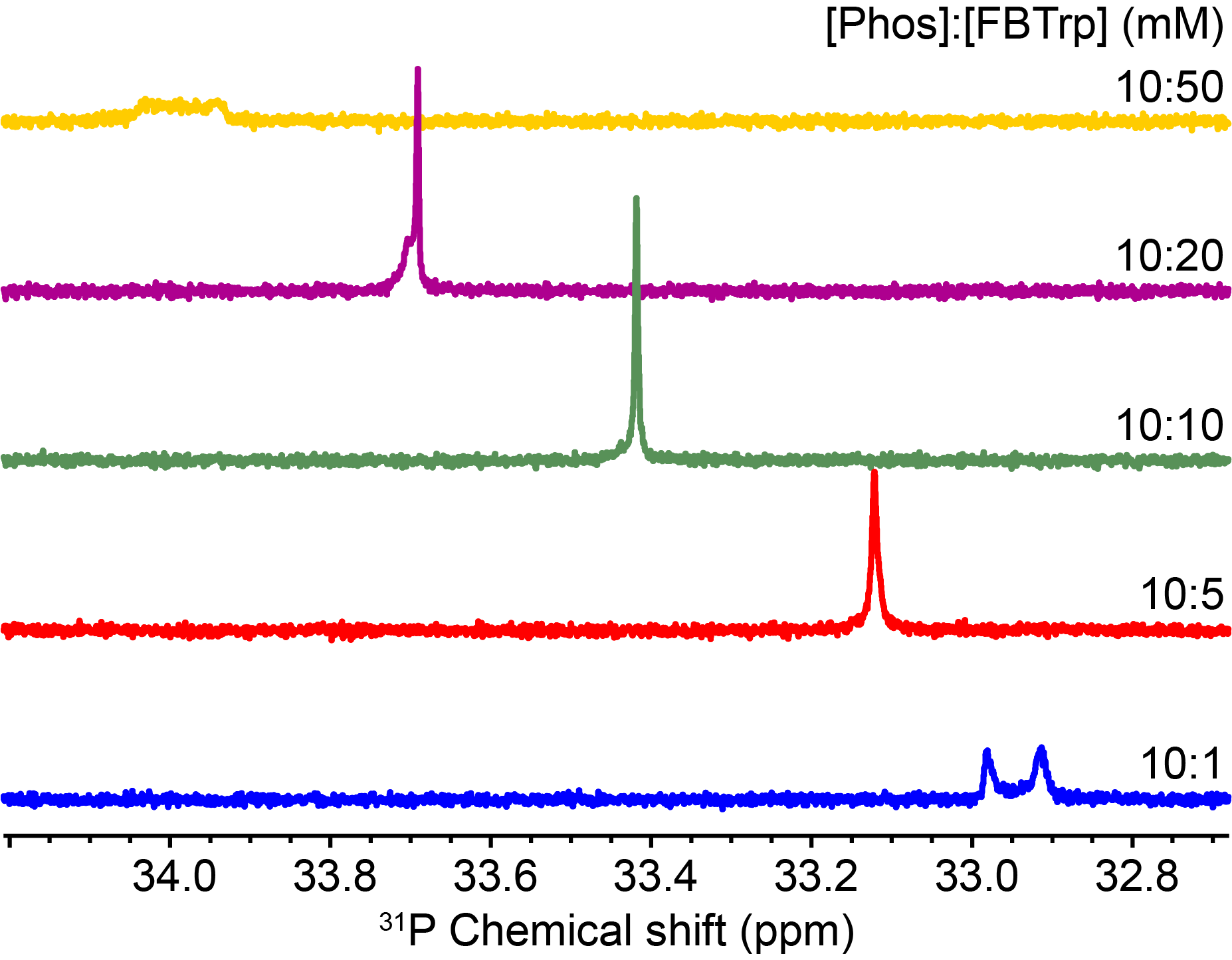}
\caption{\phos\ spectra at 20 T and 298 K of samples containing ethylated phosphorus chiral probe (\DEprobe) and \fmocS\ at several ratios. The concentration of chiral probe was maintained constant at 10 mM while concentrations of \fmocS\ varied from 1 mM to 50 mM. Samples were prepared by mixing stock solutions of \DEprobe\ and \fmocS\ in chloroform directly in 5 mm NMR tubes and were allowed to equilibrate at room temperature for at least 1 hr. }
\label{fgr:SIEthylPhosRatioPhos}
\end{figure*}

\begin{figure*}
\centering
\includegraphics[width=3.3in]{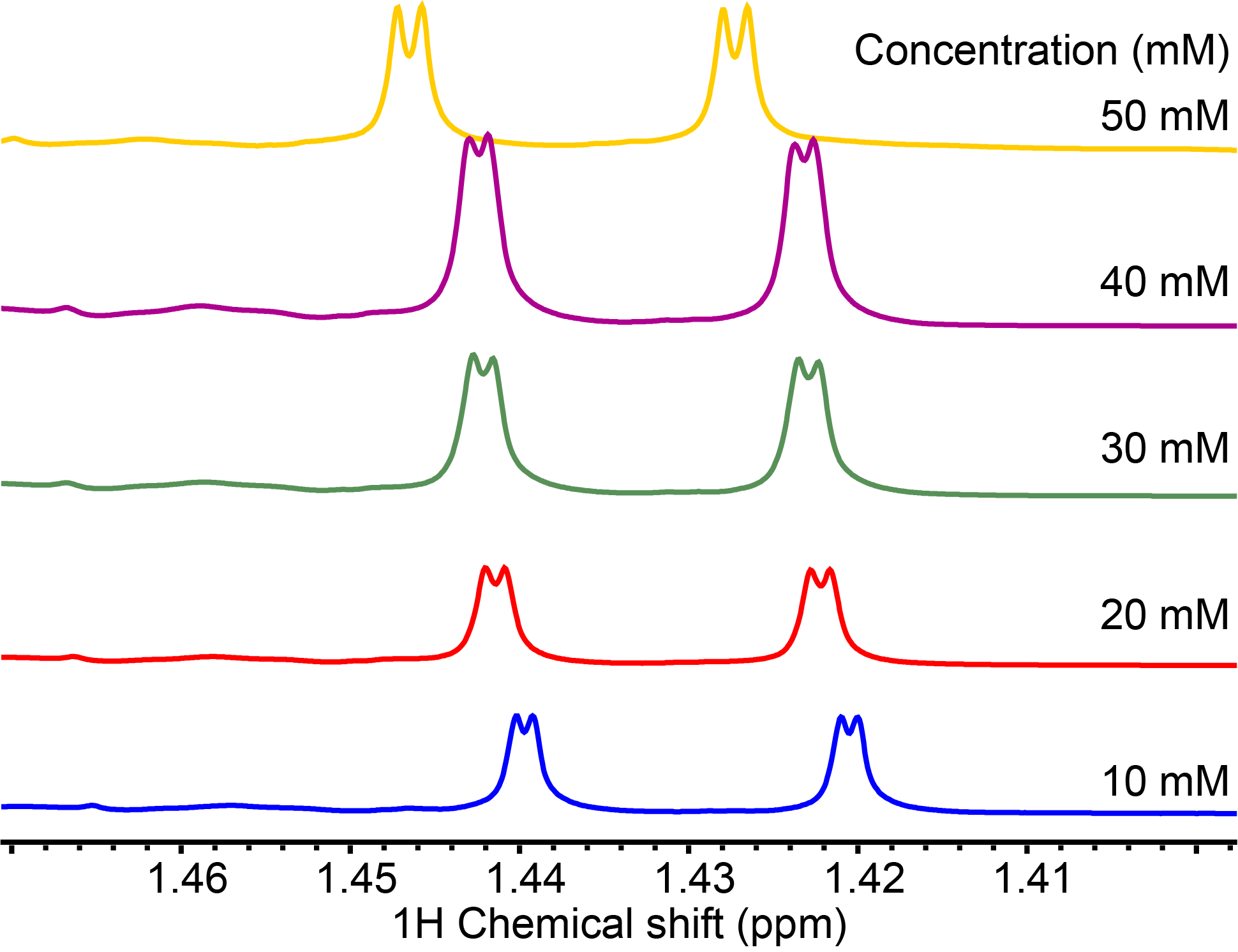}
\caption{\proton\ NMR spectra showing the effect of increasing the concentration of \fmocS\ and \DEprobe\ in tandem at a 1:1 ratio in chloroform-d. Overall small increase in \deltad\ is observed as the concentration is increased.}
\label{fgr:SIEthylPhosConcProton}
\end{figure*}

\begin{figure*}
\centering
\includegraphics[width=3.3in]{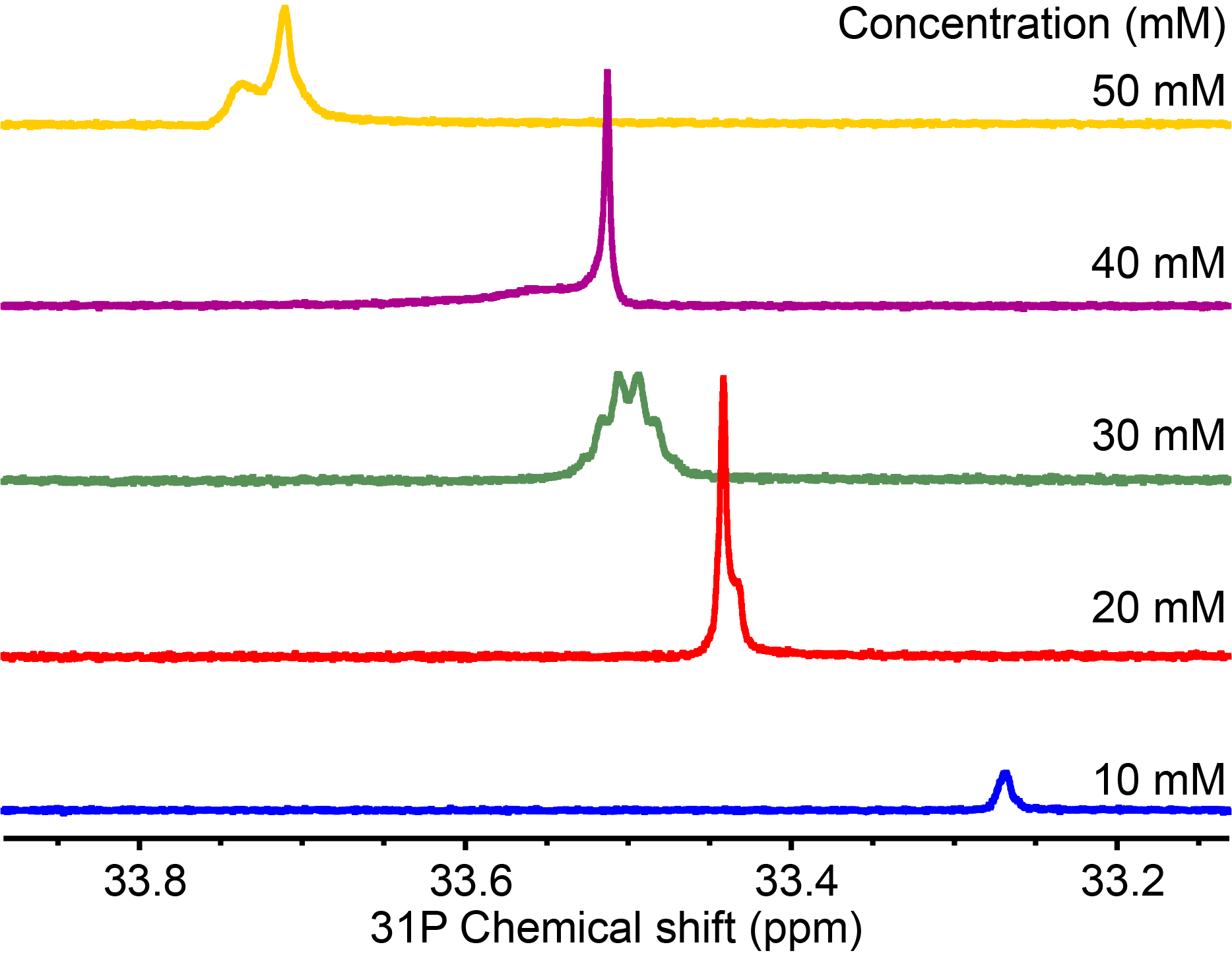}
\caption{\phos\ NMR spectra showing the effect of increasing the concentration of \fmocS\ in tandem with \DEprobe\ at a 1:1 ratio in chloroform-d . Overall a shift towards higher frequency is observed, similar to that seen in \probe\ under the same conditions. None of the tested conditions were able to produce diastereomeric splitting in \phos\ signals with \DEprobe.}
\label{fgr:SIEthylPhosConcPhos}
\end{figure*}

\begin{table*}
\caption{Relative energies $\Delta E_i$ with thermodynamic corrections of conformers of cyclohexyl \textit{P}-methylphosphonamidate at the level of PBE0/def2-TZVPP and corresponding Boltzmann weights $w_i = \mathrm{exp}(-\Delta E_i/k_\mathrm{B} T)/[\sum_j \mathrm{exp}(-\Delta E_j/k_\mathrm{B} T)]$ at $T=298.15$ K and 1 hPa pressure. NMR parameters computed at the level of 2c-ZORA-BHandH/[dyall.aae3z+sp(P,N,O,CH$_3$);IGLO-III] as described in the computational details section.}
\begin{tabular}{rrrrrrrrrrrr}
\hline
            Conformer& $\Delta E/(\mathrm{kJ/mol})$ & $w_i$ & $^{1}J_{\mathrm{^{31}P-^{13}C}}$/Hz & \multicolumn{3}{c}{$^{2}J_{\mathrm{^{31}P-^{1}H}}$/Hz} & $\delta_{\mathrm{^{31}P}}$/ppm & $\Delta_\mathrm{PV}\nu_{\mathrm{^{31}P}}/\mathrm{\upmu Hz}$ &   \multicolumn{3}{c}{$\Delta_\mathrm{PV}\nu_{\mathrm{^{1}H}}/\mathrm{nHz}$} \\
\hline           
                   1 &$   0.2 $&     $    0.324404 $& $135.0 $& $-14.0 $& $-11.1 $& $-12.3 $& $ 25.7 $&$ -0.68$ & $ 1.20$ & $ 0.38$ & $-0.75$ \\
                   2 &$   0.8 $&     $    0.257766 $& $122.7 $& $ -9.8 $& $-15.7 $& $ -9.7 $& $ 22.2 $&$ -0.59$ & $-0.52$ & $ 0.89$ & $ 0.49$ \\
                   3 &$   1.9 $&     $    0.166732 $& $133.7 $& $-13.2 $& $-11.0 $& $-12.0 $& $ 25.1 $&$ -0.71$ & $ 1.14$ & $ 0.14$ & $-0.60$ \\
                   4 &$   2.2 $&     $    0.146540 $& $121.5 $& $ -9.9 $& $-10.1 $& $-14.9 $& $ 21.4 $&$ -0.59$ & $ 0.21$ & $-0.45$ & $ 0.89$ \\
                   5 &$   3.5 $&     $    0.087793 $& $124.7 $& $-10.0 $& $-11.5 $& $-12.5 $& $ 28.0 $&$ -1.17$ & $ 0.19$ & $ 1.41$ & $-1.84$ \\
                   6 &$   9.1 $&     $    0.009170 $& $138.3 $& $-15.0 $& $-12.3 $& $-11.3 $& $ 26.9 $&$ -0.58$ & $ 1.58$ & $ 0.36$ & $-1.19$ \\
                   7 &$  10.3 $&     $    0.005583 $& $126.6 $& $ -9.0 $& $-16.4 $& $-10.9 $& $ 22.1 $&$ -0.38$ & $-0.93$ & $ 1.01$ & $ 0.77$ \\
                   8 &$  14.1 $&     $    0.001205 $& $127.0 $& $-10.4 $& $ -8.8 $& $-16.3 $& $ 19.1 $&$ -1.12$ & $ 0.62$ & $-0.17$ & $ 1.08$ \\
                   9 &$  16.5 $&     $    0.000447 $& $115.4 $& $ -9.0 $& $-16.0 $& $ -8.3 $& $ 22.4 $&$  0.56$ & $-1.46$ & $-1.21$ & $ 2.31$ \\
                  10 &$  18.0 $&     $    0.000248 $& $122.9 $& $-12.0 $& $-15.0 $& $ -7.9 $& $ 18.8 $&$ -0.74$ & $-3.00$ & $ 0.63$ & $ 2.78$ \\
                  11 &$  24.0 $&     $    0.000022 $& $134.0 $& $-10.9 $& $-12.2 $& $-13.7 $& $ 25.3 $&$ -0.68$ & $ 0.59$ & $-0.81$ & $ 1.11$ \\
                  12 &$  25.3 $&     $    0.000013 $& $122.0 $& $-10.1 $& $ -8.7 $& $-15.2 $& $ 20.9 $&$ -0.40$ & $ 1.35$ & $-1.52$ & $ 1.32$ \\
                  13 &$  25.3 $&     $    0.000013 $& $133.3 $& $-12.0 $& $-13.1 $& $-10.9 $& $ 24.2 $&$ -0.86$ & $-0.18$ & $ 0.90$ & $ 0.04$ \\
                  14 &$  25.4 $&     $    0.000013 $& $134.0 $& $-12.3 $& $-13.8 $& $-10.9 $& $ 25.3 $&$ -1.00$ & $-0.67$ & $ 1.06$ & $ 0.12$ \\
                  15 &$  25.6 $&     $    0.000011 $& $122.5 $& $-15.2 $& $ -9.5 $& $-10.0 $& $ 21.8 $&$ -0.54$ & $ 0.80$ & $ 0.74$ & $-0.67$ \\
                  16 &$  25.6 $&     $    0.000012 $& $122.8 $& $-15.6 $& $ -9.6 $& $ -9.7 $& $ 22.0 $&$ -0.55$ & $ 0.91$ & $ 0.71$ & $-0.61$ \\
                  17 &$  26.0 $&     $    0.000010 $& $135.0 $& $-12.4 $& $-14.1 $& $-11.2 $& $ 25.5 $&$ -0.80$ & $-0.78$ & $ 1.14$ & $ 0.33$ \\
                  18 &$  26.3 $&     $    0.000009 $& $124.8 $& $-16.1 $& $-10.2 $& $-10.3 $& $ 21.9 $&$ -0.70$ & $ 0.82$ & $ 0.43$ & $-0.48$ \\
                  19 &$  26.9 $&     $    0.000007 $& $136.1 $& $-14.4 $& $-11.5 $& $-12.5 $& $ 25.5 $&$ -0.64$ & $ 1.21$ & $ 0.57$ & $-0.79$ \\
                  20 &$  33.7 $&     $    0.000000 $& $113.3 $& $-16.9 $& $ -6.4 $& $ -9.1 $& $ 18.0 $&$  0.25$ & $-2.20$ & $ 1.11$ & $ 0.10$ \\
                  21 &$  34.4 $&     $    0.000000 $& $127.7 $& $ -8.6 $& $-16.5 $& $-10.9 $& $ 22.3 $&$ -0.23$ & $-0.93$ & $ 1.30$ & $ 0.46$ \\
                  22 &$  35.2 $&     $    0.000000 $& $113.0 $& $-13.3 $& $ -7.6 $& $-12.1 $& $ 17.9 $&$  0.09$ & $-2.15$ & $ 1.97$ & $ 1.16$ \\
                  23 &$  36.3 $&     $    0.000000 $& $118.2 $& $-12.0 $& $ -7.0 $& $-15.0 $& $ 15.9 $&$ -1.19$ & $ 0.07$ & $ 1.11$ & $-0.41$ \\
\hline                
                \multicolumn{3}{c}{Boltzmann weighted average}     & $128.7$ &$ -11.8$ &$ -12.2$ &$ -12.0$ & $ 24.2$ &  $ -0.69$ & $ 0.50$ &  $0.44$ & $-0.25 $ \\                
\hline                
\end{tabular}
\end{table*}



\balance


\bibliography{rsc} 
\bibliographystyle{rsc} 